\documentclass[reprint,superscriptaddress,amappendixath,amssymb,aps]{revtex4-2}
\usepackage{graphicx}
\usepackage{dcolumn}
\usepackage{bm}
\usepackage{amsmath}
\usepackage{qcircuit}
\usepackage{algpseudocode}
\usepackage{algorithm}
\usepackage{subfigure}
\usepackage[pass]{geometry}
\usepackage{float}
\makeatletter
\renewcommand*\env@matrix[1][\arraystretch]{%
\edef\arraystretch{#1}%
\hskip -\arraycolsep
\let\@ifnextchar\new@ifnextchar
\array{*\c@MaxMatrixCols c}}
\makeatother
\begin{document}
\preprint{APS/123-QED}
\title{Boson sampling enhanced quantum chemistry}
\author{Zhong-Xia Shang}
\affiliation{Hefei National Research Center for Physical Sciences at the Microscale and School of Physical Sciences,
University of Science and Technology of China, Hefei 230026, China}
\affiliation{Shanghai Research Center for Quantum Science and CAS Center for Excellence in Quantum Information and Quantum Physics,
University of Science and Technology of China, Shanghai 201315, China}
\affiliation{Hefei National Laboratory, University of Science and Technology of China, Hefei 230088, China}
\author{Han-Sen Zhong}
\affiliation{Shanghai Artificial Intelligence Laboratory, Shanghai, 200232, China}
\author{Yu-Kun Zhang}
\affiliation{School of Computer Science, Peking University, Beijing 100871, China}
\affiliation{Center on Frontiers of Computing Studies, Peking University, Beijing 100871, China}
\author{Cheng-Cheng Yu}
\affiliation{Hefei National Research Center for Physical Sciences at the Microscale and School of Physical Sciences,
University of Science and Technology of China, Hefei 230026, China}
\affiliation{Shanghai Research Center for Quantum Science and CAS Center for Excellence in Quantum Information and Quantum Physics,
University of Science and Technology of China, Shanghai 201315, China}
\affiliation{Hefei National Laboratory, University of Science and Technology of China, Hefei 230088, China}
\author{Xiao Yuan}
\affiliation{Center on Frontiers of Computing Studies, Peking University, Beijing 100871, China}
\affiliation{School of Computer Science, Peking University, Beijing 100871, China}
\author{Chao-Yang Lu}
\affiliation{Hefei National Research Center for Physical Sciences at the Microscale and School of Physical Sciences,
University of Science and Technology of China, Hefei 230026, China}
\affiliation{Shanghai Research Center for Quantum Science and CAS Center for Excellence in Quantum Information and Quantum Physics,
University of Science and Technology of China, Shanghai 201315, China}
\affiliation{Hefei National Laboratory, University of Science and Technology of China, Hefei 230088, China}
\author{Jian-Wei Pan}
\affiliation{Hefei National Research Center for Physical Sciences at the Microscale and School of Physical Sciences,
University of Science and Technology of China, Hefei 230026, China}
\affiliation{Shanghai Research Center for Quantum Science and CAS Center for Excellence in Quantum Information and Quantum Physics,
University of Science and Technology of China, Shanghai 201315, China}
\affiliation{Hefei National Laboratory, University of Science and Technology of China, Hefei 230088, China}
\author{Ming-Cheng Chen}
\affiliation{Hefei National Research Center for Physical Sciences at the Microscale and School of Physical Sciences,
University of Science and Technology of China, Hefei 230026, China}
\affiliation{Shanghai Research Center for Quantum Science and CAS Center for Excellence in Quantum Information and Quantum Physics,
University of Science and Technology of China, Shanghai 201315, China}
\affiliation{Hefei National Laboratory, University of Science and Technology of China, Hefei 230088, China}
\begin{abstract}
In this work, we give a hybrid quantum-classical algorithm for solving electronic structure problems of molecules using only linear quantum optical systems. The variational ansatz we proposed is a hybrid of non-interacting Boson dynamics and classical computational chemistry methods, specifically, the Hartree-Fock method and the Configuration Interaction method. The Boson part is built by a linear optical interferometer which is easier to realize compared with the well-known Unitary Coupled Cluster (UCC) ansatz composed of quantum gates in conventional VQE and the classical part is merely classical processing acting on the Hamiltonian. We called such ansatzes Boson Sampling-Classic (BS-C). The appearance of permanents in the Boson part has its physical intuition to provide different kinds of resources from commonly used single-, double-, and higher-excitations in classical methods and the UCC ansatz to exploring chemical quantum states. Such resources can help enhance the accuracy of methods used in the classical parts. We give a scalable hybrid homodyne and photon number measurement procedure for evaluating the energy value which has intrinsic abilities to mitigate photon loss errors and discuss the extra measurement cost induced by the no Pauli exclusion principle for Bosons with its solutions. To demonstrate our proposal, we run numerical experiments on several molecules and obtain their potential energy curves reaching chemical accuracy.
\end{abstract}

\maketitle
\section{Introduction}
Linear quantum optical systems (LQOS) \cite{kok2007linear} have the advantage that the photons are very clean and more robust to the noise compared with other architectures. Therefore, LQOS have played important roles in the development of quantum information science from the first quantum teleportation experiment \cite{bouwmeester1997experimental} to recent demonstrations of quantum advantage \cite{zhong2020quantum,zhong2021phase,madsen2022quantum}. However, this advantage comes with the lack of natural interactions between photons as a price which can create non-linearity that is essential for universal quantum computing. To enable universal quantum computing on LQOS, one can introduce non-linearity by feedforward measurements on ancilla qubits such as the KLM proposal \cite{knill2001scheme} and Measurement-based quantum computing \cite{raussendorf2003measurement}, which however, is hard to implement in reality and has made LQOS fall behind ion traps \cite{cirac1995quantum} and superconducting circuits \cite{blais2004cavity}.
\begin{figure*}[htb]
\centering
\includegraphics[width=1\textwidth]{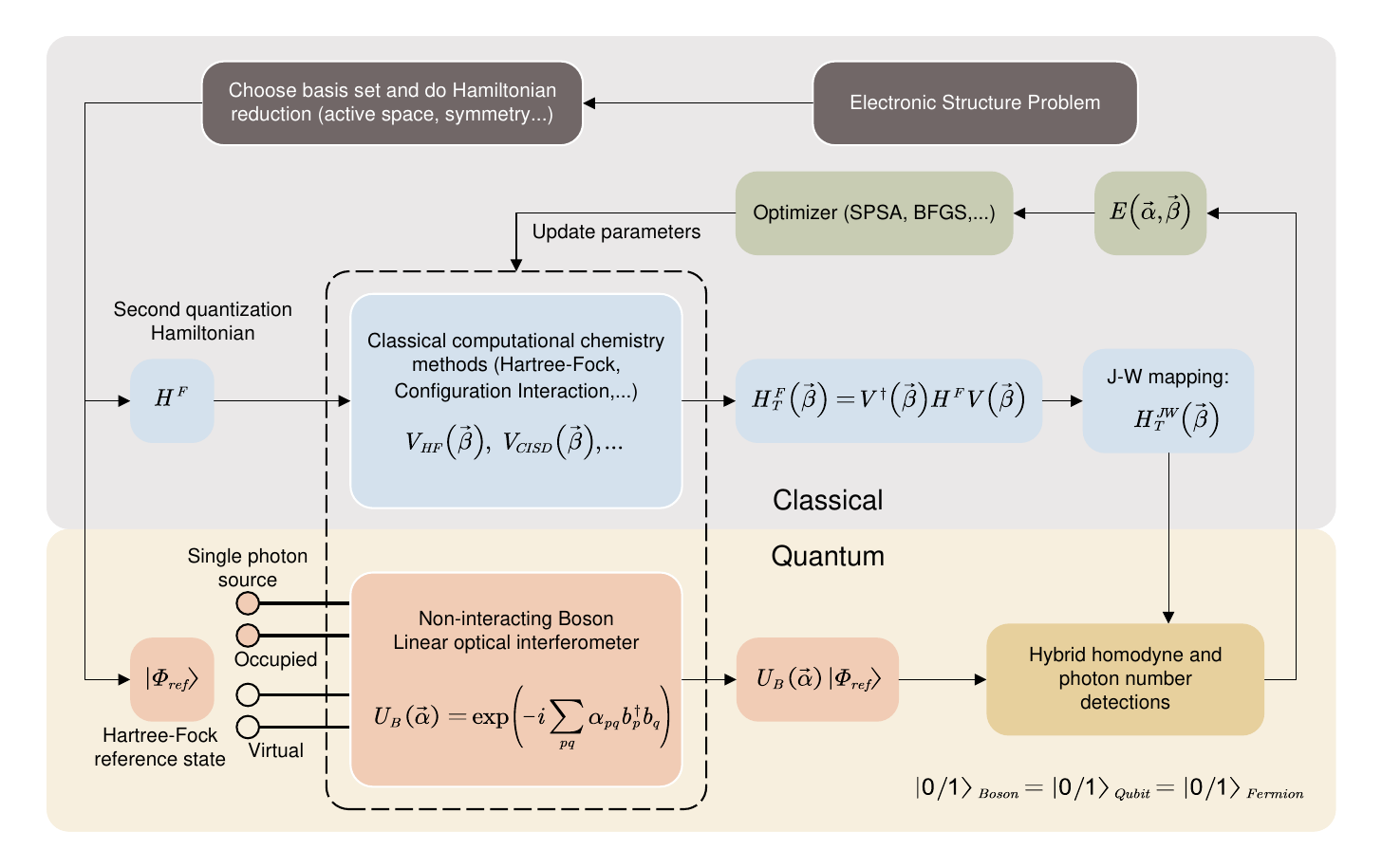}
\caption{The framework of BS-C VQE. In BS-C VQE encoding, the 0, 1 occupation of Boson, the 0, 1 of the qubit, and the 0, 1 occupation of Fermion are equivalent. Given the electronic structure Hamiltonian of a molecule, the task is to obtain its ground energy. Typically, we choose a suitable basis set and do Hamiltonian reduction if necessary to obtain the second quantization Hamiltonian $H^F$ and the HF reference state $\vert\varPhi_{ref} \rangle$. In BS-C VQE, $H^F$ will be transformed by classical operations like HF and CISD. The transformed Hamiltonian is then mapped to qubit Hamiltonian by JW mapping. The reference state gives the information on the initial photon source preparation. The occupied orbitals correspond to optical modes with single-photon input and the virtual orbitals correspond to modes with vacuum input. Such an initial state is then passed to parametrized non-interacting Boson dynamics i.e. LOI. The output state is used for the hybrid measurements to obtain energy expectation value which as the cost function is fed to the classical optimizer to update the parameters $\vec{\alpha}$ and $\vec{\beta}$ for the next iteration until converged.}
\label{p1}
\end{figure*}

Passive LQOS i.e. LQOS without feedforward measurements could not be used to make a universal quantum computer is a consensus. But can it show advantages on certain tasks that are classically hard? Aaronson and Arkhipov \cite{aaronson2011computational} gave a yes answer, which is quite a surprise to the community. Specifically, they showed sampling the outputs of a Linear Optical Interferometer (LOI) with Fock state inputs can not likely be simulated by a classical computer known as the Boson Sampling (BS) problem. Due to its friendliness to the experiments, the BS problem has made passive LQOS a promising platform to show early-stage quantum advantages. This milestone was first finished by Zhong et al \cite{zhong2020quantum} recently using Gaussian Boson Sampling, a variant of BS, and was improved by later works \cite{zhong2021phase,madsen2022quantum,deng2023gaussian}. However, the demonstration of quantum advantage won't change the fact that passive LQOS is non-universal. Thus, the next outstanding challenge is to find specific problems of practical interest that passive LQOS can solve and have potential advantages. While there have been several proposals on molecular docking \cite{banchi2020molecular}, molecular vibrational spectra \cite{huh2015boson} and graph theory \cite{arrazola2018using}, further efforts are still required for more applications.

When we look at other platforms in the Noisy Intermediate-Scale Quantum (NISQ) era \cite{preskill2018quantum}, solving Electronic Structure Problem (ESP) \cite{mcardle2020quantum} using the Variational Quantum Eigensolver (VQE) \cite{mcclean2016theory,cerezo2021variational} is one of the most anticipated applications. For ESP, we aim to solve the ground energy of the corresponding chemical Hamiltonian to help predict the chemical reaction rates. This can be done by VQE where a classical optimizer is used to optimize the energy expectation measured from the output of a shallow quantum circuit ansatz. The workflow of VQE has some degree of resilience to the noise, which makes solving ESP using NISQ quantum computers possible. Since solving ESP exactly requires exponential resources (Full Configuration Interaction (FCI)) using classical computers, quantum computing may manifest potential advantages. However, VQE is not friendly to passive LQOS since there is no natural correspondence from multi-mode Boson Fock space to multi-qubit space and the quantum circuit language can not be used to describe general operations of linear interferometers.

In this work, we give a passive LQOS-based VQE algorithm for solving ESP. The circuit ansatz in this algorithm is a hybrid of non-interacting Boson dynamics and classical computational chemistry methods. In the following, we will introduce this algorithm from its physical intuition to its workflow and implementation. Specifically, we will talk about how to measure the cost function efficiently by a combination of homodyne and photon number detections. We will also present several numerical results to show the properties and the performance of this algorithm. 

\section{Background}
In quantum chemistry \cite{mcardle2020quantum}, we are interested in the electronic structures of molecules. Under the Born-Oppenheimer approximation, treating the nuclei as classical point charges, the Hamiltonian of a molecule only has the kinetic energy terms of the electrons, the electrons' Coulomb interaction with the nuclei, and the electron-electron Coulomb repulsion left. For ESP, we aim to find the ground energy of this electronic structure Hamiltonian. However, this Hamiltonian is currently in the first quantization picture and we have to take care of the Pauli exclusion principle explicitly. To simplify the situation, we typically take the Hamiltonian to the second quantization picture where electrons can be created or annihilated from a set of orbitals (e.g. the STO-3G basis set used in our numerical examples):
\begin{equation}\label{e1}
H^F=\sum_{pq}h_{pq}f_p^\dag f_q+\frac{1}{2}\sum_{pqrs}h_{pqrs}f_p^\dag f_q^\dag f_r f_s
\end{equation}
where $f_p^\dag$ and $f_p$ are the Fermion creation and annihilation operations respectively and the $h_{pq}$ and the $h_{pqrs}$ correspond to the one- and two-electron integrals. The form of the Hamiltonian $H^F$ naturally preserves the number of electrons. If we consider $M$ orbitals as the basis set and the molecule contains $N$ electrons, then solving ESP corresponds to finding the smallest eigenenergy in the $N$-electron subspace of dimension $\binom{M}{N}$ which we will denote as the legal cluster. 
\begin{figure*}[htb]
\centering
\includegraphics[width=1\textwidth]{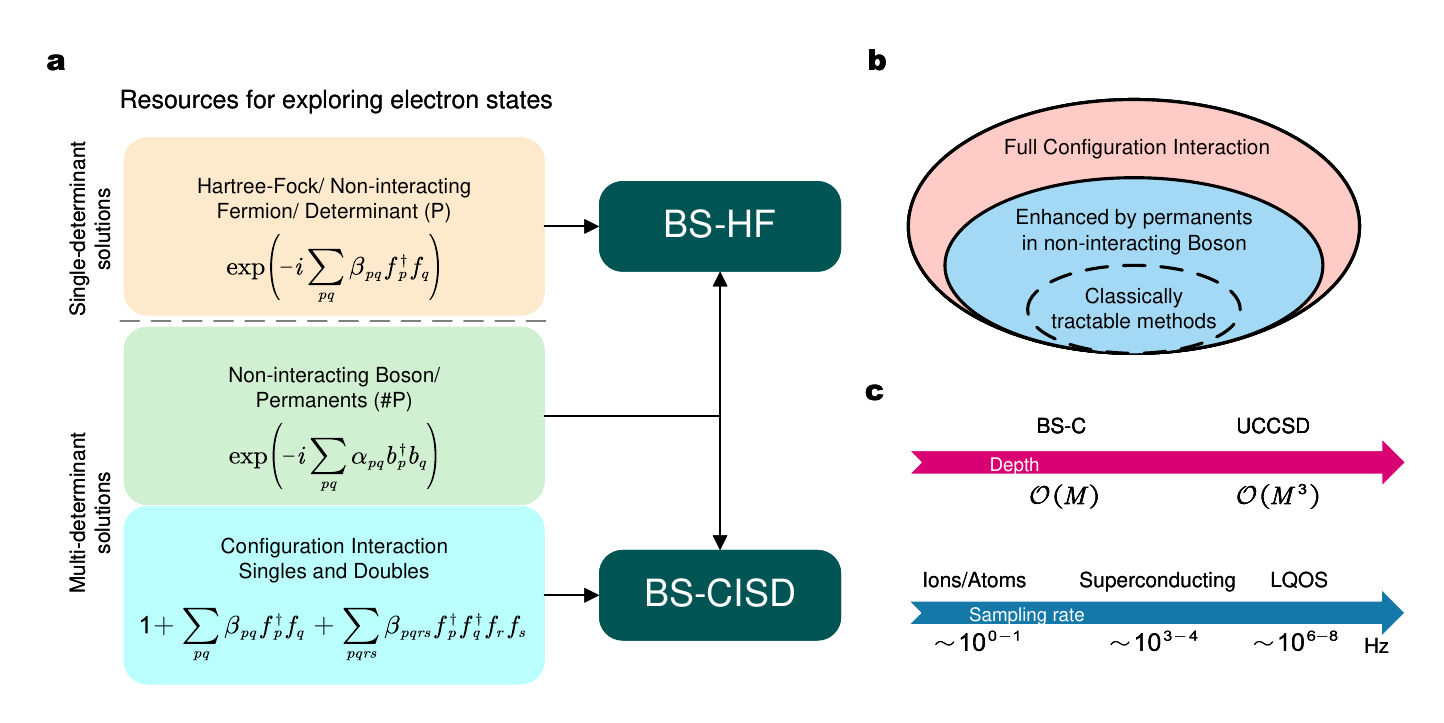}
\caption{Properties of BS-C ansatz. $\textbf{a-b}$: In the Schrödinger picture, we can see both the non-interacting Boson dynamics and the classical operations are acted on the reference state. Different operations give different resources to explore the legal cluster. UCCSD and classical methods like HF and CISD give resources for electron excitations with various orders. In contrast, permanents appearing in non-interacting Boson dynamics is a different type of resource that results from the differences between Bosons and Fermions. Also, since permanent is classically intractable, it is a quantum resource that may manifest potential advantages. Thus, the main idea of BS-C is to use such quantum resources to enhance the accuracy of classical methods. When the classical method is HF, we have BS-HF. When the classical method is CISD, we have BS-CISD. Note that the classical methods are not restricted to these two. $\textbf{c}$: BS-C has the attractive advantage of requiring much shallower depth than UCCSD. Also, the LQOS for our proposal can have a much higher sampling rate than other platforms.}
\label{p2}
\end{figure*}

The Hartree-Fock (HF) \cite{mcardle2020quantum} method is a well-known method to give an approximate solution to the problem. HF method can be understood as a VQE procedure using the non-interacting Fermion dynamics $U_F(\vec{\beta})=\exp (-i\sum_{pq}\beta_{pq}f_p^\dag f_q)$ as an ansatz to minimize the energy: $\vert HF \rangle=U_F(\vec{\beta}_{hf})\vert\phi_0\rangle,\ \vec{\beta}_{hf}=\text{argmin}_{\vec{\beta}}\langle\phi_0\vert U_F^\dag(\vec{\beta})H^F U_F(\vec{\beta}) \vert\phi_0\rangle$ where $\vert\phi_0\rangle$ is an initial product state with $N$ electrons. $U_F$ belongs to the Fermion Gaussian operations that are classically tractable due to their equivalence with the match gate circuits\cite{jozsa2008matchgates}. Since $U_F$ only works as a basis rotation of orbitals and contains no electron-electron interactions, the HF method can only give a single-determinant solution which is insufficient to accurately characterize the ground state, especially
in the dissociation area. 

To get more accurate results, one typically uses $\vert HF \rangle$ as the reference state $\vert \varPhi_{ref} \rangle$ for post-HF methods such as Configuration Interaction (CI) \cite{mcardle2020quantum} and Coupled Cluster (CC) \cite{mcardle2020quantum} methods to give multi-determinant solutions \cite{brandas2011advances}. Specifically, CI wavefunction is $(1+T)\vert \varPhi_{ref} \rangle$ and CC wave function is $e^T\vert \varPhi_{ref} \rangle$ with $T=T_1+T_2+...$ contains single, double, and higher order excitations. When only considering $T=T_1+T_2=\sum_{pq}\beta_{pq}f_p^\dag f_q+\sum_{pqrs}\beta_{pqrs}f_p^\dag f_q^\dag f_r f_s$, we have the CI singles and doubles (CISD) and CC singles and doubles (CCSD). For quantum computing, the famous Unitary Coupled Cluster (UCC) ansatz \cite{romero2018strategies,anand2022quantum} used for VQE is the unitary version of CCSD whose dynamics can be expressed as $e^{T-T^\dag}$. To implement commonly used UCC Singles and Doubles (UCCSD) on quantum computers, we usually use the Trotter decomposition to approximate the full dynamics in terms of local unitary operations, which gives rise to deep circuits intractable in NISQ devices. The idea of these methods is to introduce additional resources of interacting Fermion terms to explore states beyond a single determinant in the legal cluster. As the expressivity power of the ansatz increases by including more degrees of
freedom, a higher accuracy may be achieved.

Let us briefly introduce BS and make a comparison with Fermion's version. The general operations of a multi-mode LOI are non-interacting Boson dynamics that can be expressed as $U_B(\vec{\alpha})=\exp(-i\sum_{pq}\alpha_{pq} b_p^\dag b_q)$ where $b_p^\dag$ and $b_p$ are the Boson creation and annihilation operations respectively. The (anti-)commutation relations of $b^\dag$ and $b$ are different from $f^\dag$ and $f$, which reflects the different behaviors between Bosons and Fermions. This also leads to the difference in the complexity between the BS and the Fermion Sampling (FS) \cite{brod2021bosons}. More specifically, the transition amplitude from an initial product state $\vert S\rangle=\vert s_1s_2...s_M\rangle$ to an output product state $\vert T\rangle=\vert t_1t_2...t_M\rangle$ ($s_p$ and $t_p$ are natural numbers for Bosons and 0, 1 for Fermions.) under the non-interacting Boson and the Fermion dynamics respectively can be expressed as:
\begin{align}\label{e3}
&\langle T\vert U_B(\vec{\alpha})\vert S\rangle=\frac{\text{Per}(U_{B,ST}(\vec{\alpha}))}{\sqrt{\Pi_p s_p!\Pi_p t_p!}}\nonumber\\&\langle T\vert U_F(\vec{\beta})\vert S\rangle=\text{Det}(U_{F,ST}(\vec{\beta}))
\end{align}
where $\text{Per}$ and $\text{Det}$ denote the matrix permanent and the matrix determinant and $U_{B,ST}(\vec{\alpha})$ ($U_{F,ST}(\vec{\beta})$) is generated from the mode transformation matrix $e^{i\alpha}$ ($e^{i\beta}$) corresponding to $U_B(\vec{\alpha})$ ($U_F(\vec{\beta})$) \cite{brod2021bosons}. Since the computation of permanent is $\textbf{\#P-hard}$ and computation of determinant are in $\textbf{P}$ complexity, we believe BS is classically intractable but FS is classically tractable \cite{aaronson2011computational}. This difference is the key inspiration for our proposal.

\section{Algorithm}
\subsection{Ansatz}
To enable passive LQOS for solving ESP, we use the single-rail encoding \cite{kok2007linear} where 0, 1 photon states $\vert0\rangle_B$ and $\vert1\rangle_B$ represents 0, 1 electron states $\vert0\rangle_F$ and $\vert1\rangle_F$. When we use the Jordan-Wigner (JW) mapping \cite{mcardle2020quantum} to transform Fermion Hamiltonians to qubit Hamiltonians, this encoding also means $\vert0\rangle_B$ and $\vert1\rangle_B$ represents qubit 0, 1 states $\vert0\rangle_Q$ and $\vert1\rangle_Q$. The conservation of photon number in passive LQOS naturally corresponds to the conservation of electron number. As mentioned above, to go beyond a single determinant and create superpositions in the legal cluster, it is important to have resources different from the HF method based on non-interacting Fermion dynamics. The comparison between BS and FS tells us that permanents in non-interacting Boson dynamics under the single-rail encoding exactly provide such a resource and are different from the high-order excitations in CI, CC, and UCC. Also, this resource is classically intractable. This observation helps us create the following ansatz which we named Boson Sampling-Classic (BS-C) for VQE:
\begin{align}\label{e4}
\vert\varPhi_{BS-C}\rangle=V_C(\vec{\beta})\vert \varPhi(\vec{\alpha})\rangle= V_C(\vec{\beta})U_B(\vec{\alpha})\vert \varPhi_{ref} \rangle
\end{align}
The $U_B(\vec{\alpha})$ part is the non-interacting Boson dynamics that serves as the additional resources to enhance the performances of classical operations $V_C(\vec{\beta})$ such as $V_{HF}(\vec{\beta})=U_B(\vec{\beta})$ in HF that leads to BS-HF and $V_{CISD}(\vec{\beta})=1+T_1+T_2$ in CISD that leads to BS-CISD. 

The realization of BS-C and the implementation of the measurement of energy require a detailed discussion. The initial single-determinant reference state $\vert \varPhi_{ref} \rangle$ directly corresponds to an initial photon state. The occupied orbitals correspond to optical modes with single photon inputs and the virtual orbitals correspond to modes with vacuum inputs. $U_B(\vec{\alpha})$ is a real part realized by a parametrized LOI whose universal realization can be found in Ref.\cite{reck1994experimental}. $V_C(\vec{\beta})$ is a virtual part that merely corresponds to classical processing acting on the Hamiltonian $H^F$. Similar ideas have recently been investigated in Ref.\cite{shang2021schr,mizukami2020orbital}. Specifically, we need to calculate the transformed Hamiltonian $H^F_T(\vec{\beta})=V_C^\dag(\vec{\beta}) H^FV_C(\vec{\beta})$ which is used to replace $H^F$ for JW transformation to get a qubit Hamiltonian $H^{JW}_T(\vec{\beta})$ for measurements. (In current encoding, $H^{JW}_T(\vec{\beta})$ has the same matrix elements as $H^F_T(\vec{\beta})$. The reason we add JW mapping is that the Pauli expression of $H^{JW}_T(\vec{\beta})$ can be easier to formulate the measurement procedure.) Calculating $H^F_T(\vec{\beta})$ is efficient for proper $V_C(\vec{\beta})$ like HF and CISD since $H^F$ in Eq.\ref{e1} is a local Fermion Hamiltonian and $V_{HF}(\vec{\beta})$ is merely a basis rotation for HF \cite{jozsa2008matchgates} and $V_{CISD}(\vec{\beta})$ is also a local operator for CISD. The resulting $H^F_T(\vec{\beta})$ is also local, which means there is only a polynomial number of Fermion terms in $H^F_T(\vec{\beta})$ and also a polynomial number of Pauli terms in $H^{JW}_T(\vec{\beta})$. The polynomial number of terms in $H^{JW}_T(\vec{\beta})$ is crucial for the measurement procedure to be scalable which will talked about later. Note that other classical blocks can also be used as the classical part as long as they satisfy the conditions on efficient measurements shown later.

\subsection{Cost function and hybrid measurement}
For the cost function, we should realize that the value $\langle \varPhi(\vec{\alpha})\vert H^{JW}_T(\vec{\beta}) \vert \varPhi(\vec{\alpha})\rangle$ doesn't correspond to the true expectation value of energy and two corrections are needed. One direct correction is inherited from $V_C(\vec{\beta})$. When $V_C(\vec{\beta})$ is non-unitary, the norm of $\vert\varPhi_{BS-C}\rangle$ will be changed. The other correction is due to the fact that $\vert \varPhi(\vec{\alpha})\rangle$ may have populations outside the encoding space since there is no Pauli exclusion principle for photons, which will also influence the length of $\vert\varPhi_{BS-C}\rangle$. Thus, the true energy value should be:
\begin{align}\label{e5}
E(\vec{\alpha},\vec{\beta})=\frac{\langle \varPhi(\vec{\alpha})\vert H^{JW}_T(\vec{\beta}) \vert \varPhi(\vec{\alpha})\rangle}{\langle \varPhi(\vec{\alpha})\vert V_C^\dag(\vec{\beta})V_C(\vec{\beta})\vert \varPhi(\vec{\alpha})\rangle}
\end{align}
We should remember that both $H^{JW}_T(\vec{\beta})$ and $V_C(\vec{\beta})$ are defined in the encoded Fermion $M$-orbital $N$-electron Hilbert space of dimension $\binom{M}{N}$ which have zero matrix element values outside this subspace. For BS-HF, $V_{HF}(\vec{\beta})$ is unitary in this subspace, $E(\vec{\alpha},\vec{\beta})$ is reduced to $\langle \varPhi(\vec{\alpha})\vert H^{JW}_T(\vec{\beta}) \vert \varPhi(\vec{\alpha})\rangle/\langle \varPhi(\vec{\alpha})\vert Q \vert \varPhi(\vec{\alpha})\rangle$ where $Q$ is the projection operator from Boson $M$-mode $N$-photon Hilbert space of dimension $\binom{M+N-1}{N}$ to the encoded Fermion $M$-orbital $N$-electron Hilbert space of dimension $\binom{M}{N}$ (Note that $\vert \varPhi(\vec{\alpha})\rangle$ is composed of only $N$-photon states). We name the value of $\langle \varPhi(\vec{\alpha})\vert Q \vert \varPhi(\vec{\alpha})\rangle$ the projection ratio. When considering the whole measurement cost of $E(\vec{\alpha},\vec{\beta})$, the value of the projection ratio is crucial which we will discuss later.

Having the definition of the cost function in Eq. \ref{e5}, the question is how to measure it. According to the transformation rule of the JW mapping, the terms we need to measure will only be composed of $I$, $Z$, $\sigma_+$, and $\sigma_-$. Under the single-rail encoding, different quantum states correspond to different photon numbers, thus, conventional photon number basis measurement in dual-encoding is not universal \cite{kok2007linear} for measuring $\sigma_-$ and $\sigma_+$. On the other hand, the homodyne measurement \cite{d2007homodyne} based on continuous variables in phase space which can be done experimentally by mixing the target quantum mode with a local oscillator by a balanced beam splitter and measuring two output modes using two photo-detectors with post-processing \cite{weedbrook2012gaussian} can only be efficient for evaluating local terms which is not compatible with the JW mapping due to the existence of $Z$ (See Appendix). Here, a term is local means it has $Z$, $\sigma_+$, and $\sigma_-$ act on a limited number of qubits. To resolve this obstacle, we give a hybrid measurement strategy where for qubits with $I$ and $Z$, we use the photon number measurements, and for qubits with $\sigma_+$, and $\sigma_-$, we use the homodyne measurements. We want to mention that the introduction of Gaussian measurements will not affect the hardness of BS \cite{chakhmakhchyan2017boson}. 

To illustrate the workflow, we can consider a term with the form $H_i= Z_1Z_2\sigma_{+,3}\sigma_{-,4}$, the value $Tr(H_i\rho)$ with $\rho$ a 4-mode optical state can be expressed as:
\begin{align}\label{e6}
&Tr(H_i\rho)=\sum_{jk}q_{jk}\langle j\vert Z_1\vert j\rangle\langle k\vert Z_2\vert k\rangle\nonumber\int_{0}^\pi\frac{d\phi_3 d\phi_4}{\pi^2}\nonumber\\&\int_{-\infty}^\infty dx_3 dx_4 p_{jk}(\phi_3,\phi_4,x_3,x_4)K(\phi_3,\phi_4,x_3,x_4,\sigma_{+,3}\sigma_{-,4})
\end{align}
where $q_{jk}p_{jk}(\phi_3,\phi_4,x_3,x_4)=\langle jk\vert\langle x_{\phi_3}x_{\phi_4}\vert\rho\vert jk\rangle\vert x_{\phi_3}x_{\phi_4}\rangle$ is the joint probability of getting $j$ photons and $k$ photons on the first two modes and obtaining the results $x_{\phi_3}$ and $x_{\phi_4}$ when measuring the quadratures $X_{3,\phi_3}$ and $X_{4,\phi_4}$ where $X_{p,\phi_p}=(a_p^\dag e^{i\phi_p}+a_p e^{-i\phi_p})/2$ and $K(\phi_1,...,\phi_M,x_1,...,x_M,\vec{\beta})$ the kernel function associated with $\sigma_{+,3}\sigma_{-,4}$ which can be efficiently calculated. Eq.\ref{e6} is a typical Monte-Carlo integration which means we can do repeated samplings to estimate the value. From the operational level, to get a sample, Eq. \ref{e6} means we can first do a photon number measurement on the first two qubits and then do a homodyne measurement on the last two qubits. For the homodyne measurement, we need to get $\phi_3$ and $\phi_4$ from a uniform distribution, and then measure the corresponding $X_{3,\phi_3}$ and $X_{4,\phi_4}$ and get the results $x_{\phi_3}$ and $x_{\phi_4}$ which are then used to calculate the value of the corresponding kernel function. 

For the measurement cost, we prove that the required number of samples following the procedure of Eq. \ref{e6} to estimate Eq. \ref{e5} is proportional to $\{2.07317^k, \epsilon^{-2},m_H,m_V,\chi^{-2}\}$ with $m_H$ the number of terms in $H^{JW}_T(\vec{\beta})$, $m_V$ the number of terms in $V_C^\dag(\vec{\beta})V_C(\vec{\beta})$ (Note that when $V_C^\dag(\vec{\beta})V_C(\vec{\beta})$ is reduced to $Q$, we can simply do pure photon number measurements to estimate the projection ratio.), $k$ the maximum number of qubits with $\sigma_-$ and $\sigma_+$ in terms, $\epsilon$ the desired precision, and $\chi$ the lower bound of the projection ratio. Note that $k$ has restricted values for HF and CISD methods.

\begin{figure*}[!htb]
\centering
\includegraphics[width=0.75\textwidth]{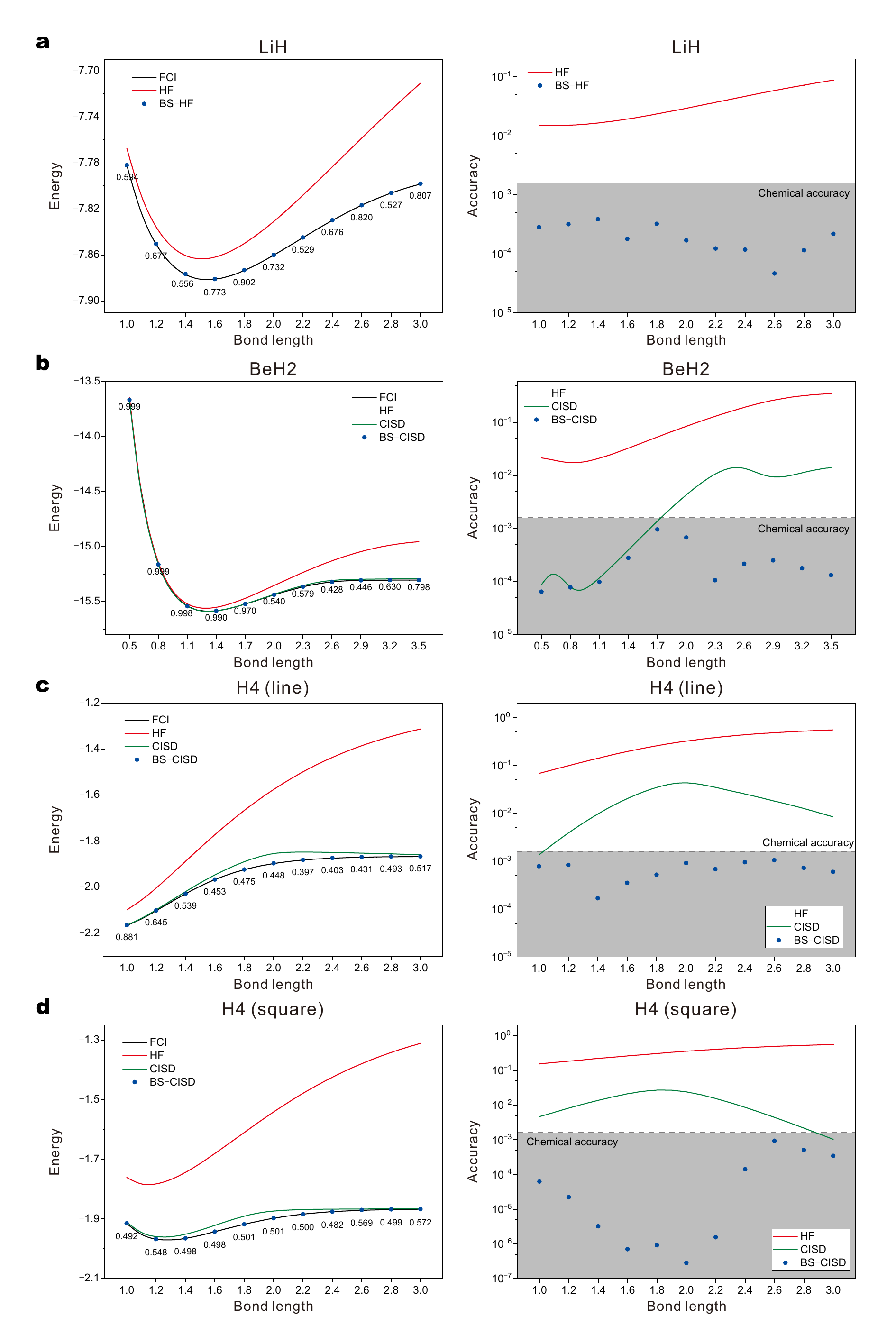}
\caption{Numerical experiments on several molecules (LiH, BeH2, and H4 with line geometry and H4 with square geometry). The potential energy of molecules as functions of bond length (Li-H bond, all Be-H bonds, all H-H bonds, and all H-H bounds) is shown on the left. Results of using BS-C, HF, CISD, and FCI methods are presented. Note that for LiH, CISD is the same as FCI. BS-HF is for LiH and BS-CISD is for the others. The numbers attached to BS-C points are the corresponding projection ratios. Absolute energy differences from FCI are shown on the right. The shaded grey region represents the area within the chemical accuracy. The classical optimizer we used is L-BFGS-B and each point of BS-C is chosen from the best result among 10 runs. The bond length unit is Angstrom and the energy unit is Hartree.}
\label{p3}
\end{figure*}

\subsection{Error mitigation against photon loss}
Currently, the above discussions treat LQOS to be ideal, which neglects the photon loss error in real devices. Photon loss as the prominent noise channel is non-negligible, especially for our proposal where photon loss will lead BS-C VQE searching outside the legal subspace. Thus, besides the corrections in Eq. \ref{e5}, we also need corrections for photon loss. Since we use a hybrid measurement strategy which seems unable to correct the photon loss error by commonly used post selections at first glance, however, we now show a procedure to correct the photon loss error whose reason can be found in the appendix. 

To illustrate the concrete correction procedure, without loss of generality, we can consider an $H_i$ only has one $\sigma_+$ and one $\sigma_-$ as an example. We have $N$ single-photon sources as input to match the correct number of electrons. First, we do the mentioned hybrid measurements. Here, since $\sigma_+\sigma_-=\vert 10\rangle\langle 01\vert$, we require to detect $N-1$ photons on $I$ and $Z$ modes with at most one photon on each mode to proceed with homodyne detections. We can do repeated such measurements to obtain a raw expectation value of $H_i$ denoted as $\langle H_i\rangle_{raw}$. Next, we need a normalization factor which can be obtained by doing repeated pure photon number measurements on all modes many times. Suppose among these measurements, we have $n_1$ times of getting $N$ photons in total on all modes with at most one photon on each mode and $n_2$ times of getting $N-1$ photons in total on $I$ and $Z$ modes with at most one photon on each mode. And among those $n_1$ events, we have $n_3$ times where there is one photon in two non-diagonal modes and $N-1$ photons in diagonal modes. The true value is then corrected by $\frac{n_3}{n_1}\frac{n_2}{n_3}\langle H_i\rangle_{raw}$. This value has already taken the projection ratio into consideration. For other $H_i$, this procedure can be easily generalized. The efficiency of this error mitigation method merely depends on the level of photon loss and has no additional non-scalable cost.

\section{Advantage discussion}
We have introduced all components of our proposal and the whole workflow of the algorithm which we named BS-C VQE has been summarized in Fig.\ref{p1}. Note that the parameters $\vec{\alpha}$ and $\vec{\beta}$ are optimized simultaneously. We also explain the physical intuitions of BS-C VQE in Fig.\ref{p2}a-b. We want to emphasize that the additional resources provided by BS are non-trivial. 

First, under our encoding, since BS and FS are different, the non-interacting Boson dynamics in LOI lead to a different transition probability distribution from non-interacting Fermion dynamics since permanents and determinants in Eq. \ref{e3} are fundamentally different. Thus, unlike HF, LOI belongs to the multi-determinant type. This ``interacting'' part can be understood as a multi-determinant reference state \cite{brandas2011advances} generator for the followed classical part and the Boson nature of these resources indicates they have small overlaps with those classical methods originated from electron excitations. Thus, when the optimization procedure is converged, we should expect to obtain more accurate ground energy estimations than using the classical methods alone. Besides, it is known that single reference methods are not capable of solving strongly correlated systems\cite{baek2023say}. Yet, the resource
consumption of classical multi-reference methods is high, which limits their practical utility. 

Second, the numerator of $E(\vec{\alpha},\vec{\beta})$ can be expanded as
\begin{align}\label{e7}
&\langle \varPhi(\vec{\alpha})\vert H^{JW}_T(\vec{\beta}) \vert \varPhi(\vec{\alpha})\rangle=\nonumber\\&\sum_{ij}H^{JW}_T(\vec{\beta})_{ij}\text{Per}(U_{B,0i}(\vec{\alpha}))\text{Per}(U_{B,0j}(\vec{\alpha})) 
\end{align}
where $0$ denotes $\vert \varPhi_{ref} \rangle$. The permanents appeared in this expression indicates the additional resources provided by BS are classically intractable, which is similar to UCCSD whose unitarity makes itself classically intractable. To have potential quantum advantages, having such resources is the basic requirement. We thus expect our method to have a certain advantage for solving strongly correlated systems. We want to mention that when compared with pure classical algorithms, performance enhancements by BS come with an additional sampling cost and the level of enhancements can vary for different molecules. If, for example, we have BS-CISD comparable with CISDT in terms of accuracy, and the sampling cost is smaller than the computational cost of CISDT, a real quantum advantage is achieved. 
\begin{figure*}[htb]
\centering
\includegraphics[width=0.8\textwidth]{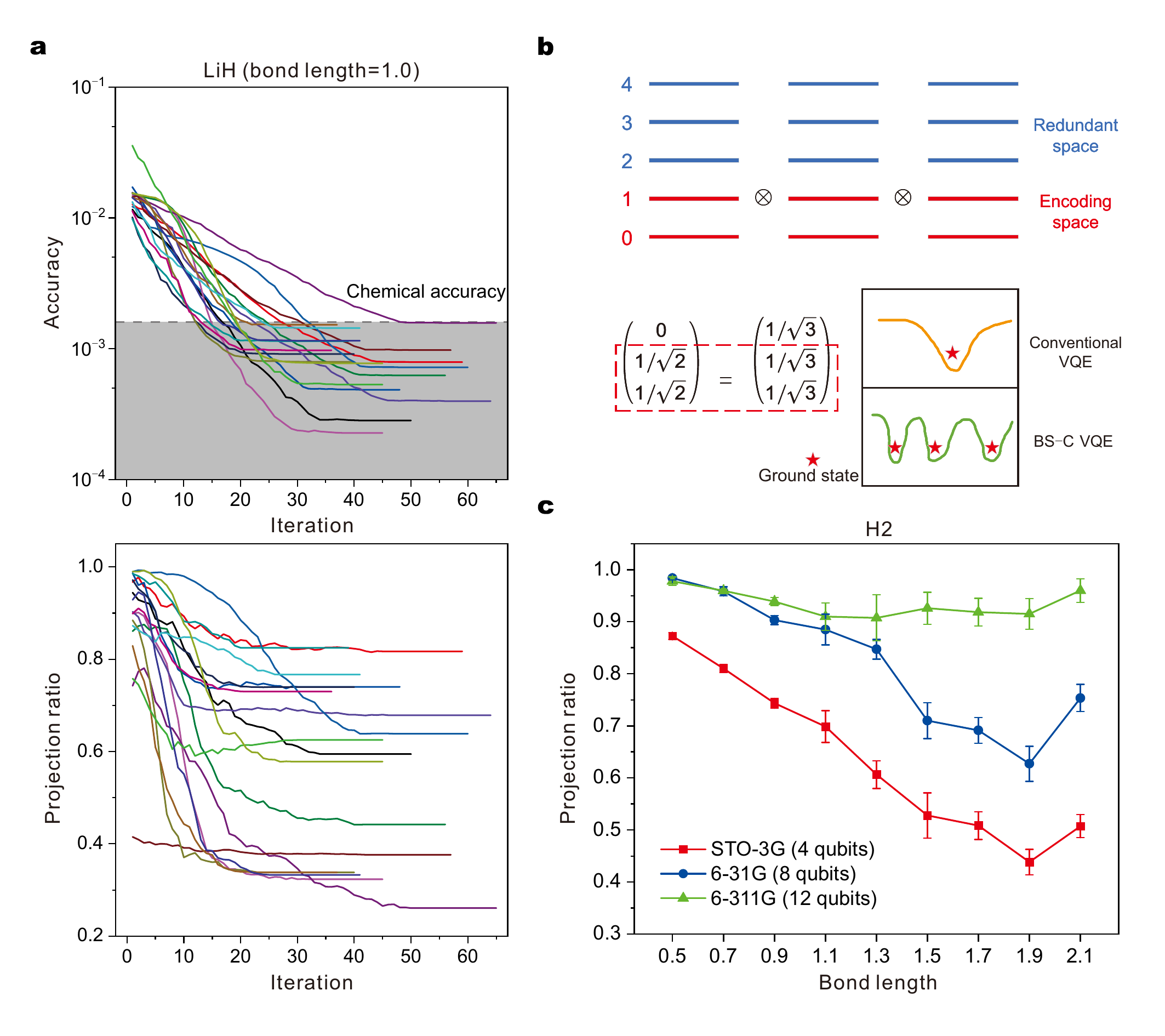}
\caption{Projection ratio of BS-C VQE. $\textbf{a}$: LiH molecule as an example. We fix the bond length as 1.0 and repeatedly run BS-C VQE. We see that all presented tests have achieved chemical accuracy. However, as explained, their corresponding projection ratios are totally different. $\textbf{b}$: In BS-C VQE encoding, we only utilize 0 and 1 occupation dimensions, and all higher photon occupations are redundant. However, after the action of LOI, part of the output state will unavoidably jump to these redundant dimensions. This leads to an interesting point of BS-C VQE that two physically different states can be treated as the same states in the encoding space. Thus, when optimizing BS-C VQE, it is possible that we can get multiple sets of parameters that correspond to the ground state. $\textbf{c}$: The number of orbitals and the number of electrons have a crucial influence on the projection ratio. The smaller of the number of electrons compared to the number of orbitals, the higher the projection ratio will appear. We test the H2 molecule with different choices of basis set to show this phenomenon. The number of electrons is 2 and the orbital number is 4 for STO-3G, 8 for 6-31G, and 12 for 6-311G. The error bars are collected from the tests that achieve chemical accuracy. We can easily see the larger number of orbitals gives higher projection ratios. We use BS-HF for LiH and H2.}
\label{p4}
\end{figure*}

When considering the realizations, BS-C VQE has its advantage over the UCCSD. Unlike UCCSD where we need a locally connected quantum circuit of depth around $\mathcal{O}(M^3)$ \cite{anand2022quantum} to build in the general case where all excitations are allowed, BS-C only needs a $\mathcal{O}(M)$-time LOI \cite{reck1994experimental} which has the same scaling as the quantum implementation of HF ansatz $U_F(\vec{\beta})$ of depth around $\mathcal{O}(M)$ \cite{google2020hartree} but is with the speed of light as the propagation speed. Thus, combined with the efficient error mitigation method we will introduce below against the photon loss, BS-C VQE is much more hardware-efficient. What's more, the LQOS for our proposal has the advantage of the sampling rate compared with other platforms, which is highly demanded for variational quantum algorithms since massive repeated measurements are required. Current state-of-the-art experiments show that the sampling rate of LQOS can reach around $10^{6-8} Hz$ \cite{wang2019boson} whereas the sampling rate is around $10^{3-4}Hz$ for superconducting qubits \cite{wu2021strong} and $10^{0-1}Hz$ for ion traps \cite{pino2021demonstration} and neutral atoms \cite{bluvstein2022quantum}. We summarize these in Fig.\ref{p2}c. 

\section{Numerical experiments}
\subsection{Solving ESP of small molecules}
We demonstrate the performance of BS-C VQE by running numerical experiments on several molecules and drawing their potential energy curves in Fig.\ref{p3}. We test molecules including LiH, BeH2, and H4 with line geometry and H4 with square geometry. All molecules use the STO-3G basis set and the active space reduction method \cite{mcardle2020quantum} is used for LiH, BeH2. The resulting JW Hamiltonians are 6, 8, 8, and 8 qubits. We add data of FCI, HF, and CISD as references. We also show the projection ratios corresponding to the BS-C VQE solutions. We can see BS-C has enhancements over their classical part and can reach the chemical accuracy ($1.6\times 10^{-3}$ Hartree) at all regions. We use BS-HF for LiH, where we can see that while we only use non-interacting Boson and non-interacting Fermion, the results are still rather accurate. The other three molecules however, are actually strongly correlated at some bond length where HF performs so poor that even BS can not give enough enhancements to reach the chemical accuracy. Thus, we use BS-CISD instead. 

\subsection{Projection ratio}
Now, we want to talk about the projection ratio. The projection ratio $\langle \varPhi(\vec{\alpha})\vert Q \vert \varPhi(\vec{\alpha})\rangle$ has great influence to the total measurement cost of estimating $E(\vec{\alpha},\vec{\beta})$. Our statistic analysis shows that the mean squared error (MSE) containing both bias and variance of estimating $E(\vec{\alpha},\vec{\beta})$ will increase by $\mathcal{O}(\langle Q\rangle^{-2})$ times compared with the ideal case with projection ratio equal to one. Thus, to make the measurements of $E(\vec{\alpha},\vec{\beta})$ truly scalable, the projection ratio can not be exponentially small, which is also the reason for the existence of sign problem \cite{loh1990sign} in quantum Monte-Carlo methods. 

We can increase the projection ratios from two aspects. The first can be understood from Fig.\ref{p4}a-b. We use BS-HF for the LiH molecule as a direct demonstration where we can see while all experiments finally give energy reached the chemical accuracy, they have totally different projection ratios. This means there are physically different states $\vert \varPhi(\vec{\alpha})\rangle$ which can be treated as the same states in the encoding space. Thus, unlike the conventional VQE where there is only one minimum energy, there may be multi-solutions in our algorithm. To help the optimization converge to those with high projection ratios, a penalty term $-\lambda \langle Q\rangle$ can be introduced to the cost function which helps solutions with higher projection ratios have lower cost function values. Another method is to only allow occupied-to-virtual photon excitations and virtual-to-virtual photon excitations to reduce the populations outside the encoding space, which is the method we used in numerical experiments. Note that this can be further adjusted by only allowing those to preserve the spin and orbital symmetries \cite{anand2022quantum}. 

These methods however are not enough since there may be situations where even the highest projection ratio among the solutions is still small. We can understand this by defining the ratio $r(M,N)=\binom{M}{N}/\binom{M+N-1}{N}$ between the dimension of Fermion Hilbert space and the dimension of Boson Hilbert space. Obviously, $r(M,N)$ is a fair judge for the projection ratio. If we fix the number of orbitals $M$ and enlarge the number of electrons $N$, $r(M,N)$ will decrease. In contrast, if we fix the number of electrons $N$ and enlarge the number of orbitals $M$, $r(M,N)$ will increase. Thus, to avoid low projection ratios, $M$ should be large enough compared to $N$. Luckily, while $N$ depends on the number of electrons in the molecule, the number of orbitals has no restrictions and can be arbitrarily large. 

In Fig.\ref{p4}c, we run BS-HF for H2 with different choices of basis sets including STO-3G, 6-31G, and 6-311G. The resulting number of qubits (spin orbitals) is 4, 8, and 12 respectively with the number of electrons fixed as 2. We can see that as the number of orbitals gets larger, the average projection ratios of the experiments that reach the chemical accuracy gets larger as well. More rigorously, we proved that if $M=\eta N^2$, $r(M,N)$ will have a lower bound $e^{1/\eta}$, which indicates the empirical requirements to make the measurements scalable. Note that such a relation between $M$ and $N$ is also a believed requirement for BS to be classically intractable \cite{lund2017quantum,aaronson2011computational}.

\section{Summary and outlook}
In summary, we give a novel hybrid quantum-classical ansatz using only passive LQOS for running VQE to solve ESP. The permanents appeared in non-interacting Boson dynamics is a new type of resource for exploring the legal cluster which can help enhance the accuracy of quantum chemistry. Thus, BS-C VQE is both hardware-efficient with only an LQI and chemical-inspired for its physical intuitions. We use BS-HF and BS-CISD as two concrete settings and show their performances by numerical results. We want to mention that the classical methods are not restricted to HF and CI. The introduction of the hybrid measurement strategy with the methods against photon loss error and low projection ratios helps to have a scalable measurement cost. Both the theory and the simulations showed that our proposal may open a road for quantum applications using passive LQOS in the NISQ era. 

Also, this proposal is not restricted to what we presented here, by the developments of LQOS, to get better accuracy, one can use entangled reference states \cite{brandas2011advances} or introduce intermediate measurements for more freedom and higher accuracy. It is worth noticing that there are still several open questions on this algorithm that need to be further investigated such as the level of performance enhancements of this algorithm on larger molecules and better measurement strategies such as developing efficient tomography methods with pure homodyne or photon number measurements \cite{olivares2019quantum} and adopting the ideas of recently developed quantum learning methods \cite{huang2020predicting}.

Finally, we briefly discuss the feasibility for an experimental demonstration. Our protocol requires a standard boson sampling (and its variants) set-up where the single-photon quantum light sources are injected into a tunable multi-mode interferometer and the outputs are resorted to hybrid homodyne and photon-number detections. All the required components have been demonstrated in previous experiments, although separately. For example, Gaussian boson sampling has been tested up to 255 output photons \cite{deng2023gaussian} and 216-mode reconfigurable optical circuits \cite{madsen2022quantum}. High-quality single-photon sources \cite{ding2016demand,senellart2017high} and high-fidelity homodyne measurements have been realized in Ref. \cite{zavatta2011high,raffaelli2018homodyne}. Therefore, our scheme is within the reach of current technologies.

We used Python packages including Qiskit \cite{cross2018ibm}, OperFermion \cite{mcclean2020openfermion}, PySCF \cite{sun2018pyscf}, and The Walrus \cite{gupt2019walrus} for parts of our simulations. The example codes can be found online \cite{shang2023bfvqe}.

More details can be found in the Appendix.

$\mathbf{Achknowledgement:}$ This work was supported by the National Natural Science Foundation of China (No. 91836303 and No. 11805197), the National Key R-D Program of China, the Chinese Academy of Sciences, the Anhui Initiative in Quantum Information Technologies, and the Science and Technology Commission of Shanghai Municipality (2019SHZDZX01). H.-S. Z. is supported by National Key R$\&$D Program of China (NO.2022ZD0160100) Shanghai Committee of Science and Technology (Grant No. 21DZ1100100). X. Y. and Y.-K. Z. are supported by the National Natural Science Foundation of China Grant (No.~12175003 and No.~12361161602),  NSAF (Grant No.~U2330201). The authors would like to thank Z-H Chen and Y-H Deng for inspiring discussions and S Aaronson and P Coveney for their fruitful comments.
\bibliography{ref.bib}

\begin{thebibliography}{50}%
\makeatletter
\providecommand \@ifxundefined [1]{%
 \@ifx{#1\undefined}
}%
\providecommand \@ifnum [1]{%
 \ifnum #1\expandafter \@firstoftwo
 \else \expandafter \@secondoftwo
 \fi
}%
\providecommand \@ifx [1]{%
 \ifx #1\expandafter \@firstoftwo
 \else \expandafter \@secondoftwo
 \fi
}%
\providecommand \natexlab [1]{#1}%
\providecommand \enquote  [1]{``#1''}%
\providecommand \bibnamefont  [1]{#1}%
\providecommand \bibfnamefont [1]{#1}%
\providecommand \citenamefont [1]{#1}%
\providecommand \href@noop [0]{\@secondoftwo}%
\providecommand \href [0]{\begingroup \@sanitize@url \@href}%
\providecommand \@href[1]{\@@startlink{#1}\@@href}%
\providecommand \@@href[1]{\endgroup#1\@@endlink}%
\providecommand \@sanitize@url [0]{\catcode `\\12\catcode `\$12\catcode `\&12\catcode `\#12\catcode `\^12\catcode `\_12\catcode `\%12\relax}%
\providecommand \@@startlink[1]{}%
\providecommand \@@endlink[0]{}%
\providecommand \url  [0]{\begingroup\@sanitize@url \@url }%
\providecommand \@url [1]{\endgroup\@href {#1}{\urlprefix }}%
\providecommand \urlprefix  [0]{URL }%
\providecommand \Eprint [0]{\href }%
\providecommand \doibase [0]{https://doi.org/}%
\providecommand \selectlanguage [0]{\@gobble}%
\providecommand \bibinfo  [0]{\@secondoftwo}%
\providecommand \bibfield  [0]{\@secondoftwo}%
\providecommand \translation [1]{[#1]}%
\providecommand \BibitemOpen [0]{}%
\providecommand \bibitemStop [0]{}%
\providecommand \bibitemNoStop [0]{.\EOS\space}%
\providecommand \EOS [0]{\spacefactor3000\relax}%
\providecommand \BibitemShut  [1]{\csname bibitem#1\endcsname}%
\let\auto@bib@innerbib\@empty
%</preamble>
\bibitem [{\citenamefont {Kok}\ \emph {et~al.}(2007)\citenamefont {Kok}, \citenamefont {Munro}, \citenamefont {Nemoto}, \citenamefont {Ralph}, \citenamefont {Dowling},\ and\ \citenamefont {Milburn}}]{kok2007linear}%
  \BibitemOpen
  \bibfield  {author} {\bibinfo {author} {\bibfnamefont {P.}~\bibnamefont {Kok}}, \bibinfo {author} {\bibfnamefont {W.~J.}\ \bibnamefont {Munro}}, \bibinfo {author} {\bibfnamefont {K.}~\bibnamefont {Nemoto}}, \bibinfo {author} {\bibfnamefont {T.~C.}\ \bibnamefont {Ralph}}, \bibinfo {author} {\bibfnamefont {J.~P.}\ \bibnamefont {Dowling}},\ and\ \bibinfo {author} {\bibfnamefont {G.~J.}\ \bibnamefont {Milburn}},\ }\bibfield  {title} {\bibinfo {title} {Linear optical quantum computing with photonic qubits},\ }\href@noop {} {\bibfield  {journal} {\bibinfo  {journal} {Reviews of modern physics}\ }\textbf {\bibinfo {volume} {79}},\ \bibinfo {pages} {135} (\bibinfo {year} {2007})}\BibitemShut {NoStop}%
\bibitem [{\citenamefont {Bouwmeester}\ \emph {et~al.}(1997)\citenamefont {Bouwmeester}, \citenamefont {Pan}, \citenamefont {Mattle}, \citenamefont {Eibl}, \citenamefont {Weinfurter},\ and\ \citenamefont {Zeilinger}}]{bouwmeester1997experimental}%
  \BibitemOpen
  \bibfield  {author} {\bibinfo {author} {\bibfnamefont {D.}~\bibnamefont {Bouwmeester}}, \bibinfo {author} {\bibfnamefont {J.-W.}\ \bibnamefont {Pan}}, \bibinfo {author} {\bibfnamefont {K.}~\bibnamefont {Mattle}}, \bibinfo {author} {\bibfnamefont {M.}~\bibnamefont {Eibl}}, \bibinfo {author} {\bibfnamefont {H.}~\bibnamefont {Weinfurter}},\ and\ \bibinfo {author} {\bibfnamefont {A.}~\bibnamefont {Zeilinger}},\ }\bibfield  {title} {\bibinfo {title} {Experimental quantum teleportation},\ }\href@noop {} {\bibfield  {journal} {\bibinfo  {journal} {Nature}\ }\textbf {\bibinfo {volume} {390}},\ \bibinfo {pages} {575} (\bibinfo {year} {1997})}\BibitemShut {NoStop}%
\bibitem [{\citenamefont {Zhong}\ \emph {et~al.}(2020)\citenamefont {Zhong}, \citenamefont {Wang}, \citenamefont {Deng}, \citenamefont {Chen}, \citenamefont {Peng}, \citenamefont {Luo}, \citenamefont {Qin}, \citenamefont {Wu}, \citenamefont {Ding}, \citenamefont {Hu} \emph {et~al.}}]{zhong2020quantum}%
  \BibitemOpen
  \bibfield  {author} {\bibinfo {author} {\bibfnamefont {H.-S.}\ \bibnamefont {Zhong}}, \bibinfo {author} {\bibfnamefont {H.}~\bibnamefont {Wang}}, \bibinfo {author} {\bibfnamefont {Y.-H.}\ \bibnamefont {Deng}}, \bibinfo {author} {\bibfnamefont {M.-C.}\ \bibnamefont {Chen}}, \bibinfo {author} {\bibfnamefont {L.-C.}\ \bibnamefont {Peng}}, \bibinfo {author} {\bibfnamefont {Y.-H.}\ \bibnamefont {Luo}}, \bibinfo {author} {\bibfnamefont {J.}~\bibnamefont {Qin}}, \bibinfo {author} {\bibfnamefont {D.}~\bibnamefont {Wu}}, \bibinfo {author} {\bibfnamefont {X.}~\bibnamefont {Ding}}, \bibinfo {author} {\bibfnamefont {Y.}~\bibnamefont {Hu}}, \emph {et~al.},\ }\bibfield  {title} {\bibinfo {title} {Quantum computational advantage using photons},\ }\href@noop {} {\bibfield  {journal} {\bibinfo  {journal} {Science}\ }\textbf {\bibinfo {volume} {370}},\ \bibinfo {pages} {1460} (\bibinfo {year} {2020})}\BibitemShut {NoStop}%
\bibitem [{\citenamefont {Zhong}\ \emph {et~al.}(2021)\citenamefont {Zhong}, \citenamefont {Deng}, \citenamefont {Qin}, \citenamefont {Wang}, \citenamefont {Chen}, \citenamefont {Peng}, \citenamefont {Luo}, \citenamefont {Wu}, \citenamefont {Gong}, \citenamefont {Su} \emph {et~al.}}]{zhong2021phase}%
  \BibitemOpen
  \bibfield  {author} {\bibinfo {author} {\bibfnamefont {H.-S.}\ \bibnamefont {Zhong}}, \bibinfo {author} {\bibfnamefont {Y.-H.}\ \bibnamefont {Deng}}, \bibinfo {author} {\bibfnamefont {J.}~\bibnamefont {Qin}}, \bibinfo {author} {\bibfnamefont {H.}~\bibnamefont {Wang}}, \bibinfo {author} {\bibfnamefont {M.-C.}\ \bibnamefont {Chen}}, \bibinfo {author} {\bibfnamefont {L.-C.}\ \bibnamefont {Peng}}, \bibinfo {author} {\bibfnamefont {Y.-H.}\ \bibnamefont {Luo}}, \bibinfo {author} {\bibfnamefont {D.}~\bibnamefont {Wu}}, \bibinfo {author} {\bibfnamefont {S.-Q.}\ \bibnamefont {Gong}}, \bibinfo {author} {\bibfnamefont {H.}~\bibnamefont {Su}}, \emph {et~al.},\ }\bibfield  {title} {\bibinfo {title} {Phase-programmable gaussian boson sampling using stimulated squeezed light},\ }\href@noop {} {\bibfield  {journal} {\bibinfo  {journal} {Physical review letters}\ }\textbf {\bibinfo {volume} {127}},\ \bibinfo {pages} {180502} (\bibinfo {year} {2021})}\BibitemShut {NoStop}%
\bibitem [{\citenamefont {Madsen}\ \emph {et~al.}(2022)\citenamefont {Madsen}, \citenamefont {Laudenbach}, \citenamefont {Askarani}, \citenamefont {Rortais}, \citenamefont {Vincent}, \citenamefont {Bulmer}, \citenamefont {Miatto}, \citenamefont {Neuhaus}, \citenamefont {Helt}, \citenamefont {Collins} \emph {et~al.}}]{madsen2022quantum}%
  \BibitemOpen
  \bibfield  {author} {\bibinfo {author} {\bibfnamefont {L.~S.}\ \bibnamefont {Madsen}}, \bibinfo {author} {\bibfnamefont {F.}~\bibnamefont {Laudenbach}}, \bibinfo {author} {\bibfnamefont {M.~F.}\ \bibnamefont {Askarani}}, \bibinfo {author} {\bibfnamefont {F.}~\bibnamefont {Rortais}}, \bibinfo {author} {\bibfnamefont {T.}~\bibnamefont {Vincent}}, \bibinfo {author} {\bibfnamefont {J.~F.}\ \bibnamefont {Bulmer}}, \bibinfo {author} {\bibfnamefont {F.~M.}\ \bibnamefont {Miatto}}, \bibinfo {author} {\bibfnamefont {L.}~\bibnamefont {Neuhaus}}, \bibinfo {author} {\bibfnamefont {L.~G.}\ \bibnamefont {Helt}}, \bibinfo {author} {\bibfnamefont {M.~J.}\ \bibnamefont {Collins}}, \emph {et~al.},\ }\bibfield  {title} {\bibinfo {title} {Quantum computational advantage with a programmable photonic processor},\ }\href@noop {} {\bibfield  {journal} {\bibinfo  {journal} {Nature}\ }\textbf {\bibinfo {volume} {606}},\ \bibinfo {pages} {75} (\bibinfo {year} {2022})}\BibitemShut {NoStop}%
\bibitem [{\citenamefont {Knill}\ \emph {et~al.}(2001)\citenamefont {Knill}, \citenamefont {Laflamme},\ and\ \citenamefont {Milburn}}]{knill2001scheme}%
  \BibitemOpen
  \bibfield  {author} {\bibinfo {author} {\bibfnamefont {E.}~\bibnamefont {Knill}}, \bibinfo {author} {\bibfnamefont {R.}~\bibnamefont {Laflamme}},\ and\ \bibinfo {author} {\bibfnamefont {G.~J.}\ \bibnamefont {Milburn}},\ }\bibfield  {title} {\bibinfo {title} {A scheme for efficient quantum computation with linear optics},\ }\href@noop {} {\bibfield  {journal} {\bibinfo  {journal} {nature}\ }\textbf {\bibinfo {volume} {409}},\ \bibinfo {pages} {46} (\bibinfo {year} {2001})}\BibitemShut {NoStop}%
\bibitem [{\citenamefont {Raussendorf}\ \emph {et~al.}(2003)\citenamefont {Raussendorf}, \citenamefont {Browne},\ and\ \citenamefont {Briegel}}]{raussendorf2003measurement}%
  \BibitemOpen
  \bibfield  {author} {\bibinfo {author} {\bibfnamefont {R.}~\bibnamefont {Raussendorf}}, \bibinfo {author} {\bibfnamefont {D.~E.}\ \bibnamefont {Browne}},\ and\ \bibinfo {author} {\bibfnamefont {H.~J.}\ \bibnamefont {Briegel}},\ }\bibfield  {title} {\bibinfo {title} {Measurement-based quantum computation on cluster states},\ }\href@noop {} {\bibfield  {journal} {\bibinfo  {journal} {Physical review A}\ }\textbf {\bibinfo {volume} {68}},\ \bibinfo {pages} {022312} (\bibinfo {year} {2003})}\BibitemShut {NoStop}%
\bibitem [{\citenamefont {Cirac}\ and\ \citenamefont {Zoller}(1995)}]{cirac1995quantum}%
  \BibitemOpen
  \bibfield  {author} {\bibinfo {author} {\bibfnamefont {J.~I.}\ \bibnamefont {Cirac}}\ and\ \bibinfo {author} {\bibfnamefont {P.}~\bibnamefont {Zoller}},\ }\bibfield  {title} {\bibinfo {title} {Quantum computations with cold trapped ions},\ }\href@noop {} {\bibfield  {journal} {\bibinfo  {journal} {Physical review letters}\ }\textbf {\bibinfo {volume} {74}},\ \bibinfo {pages} {4091} (\bibinfo {year} {1995})}\BibitemShut {NoStop}%
\bibitem [{\citenamefont {Blais}\ \emph {et~al.}(2004)\citenamefont {Blais}, \citenamefont {Huang}, \citenamefont {Wallraff}, \citenamefont {Girvin},\ and\ \citenamefont {Schoelkopf}}]{blais2004cavity}%
  \BibitemOpen
  \bibfield  {author} {\bibinfo {author} {\bibfnamefont {A.}~\bibnamefont {Blais}}, \bibinfo {author} {\bibfnamefont {R.-S.}\ \bibnamefont {Huang}}, \bibinfo {author} {\bibfnamefont {A.}~\bibnamefont {Wallraff}}, \bibinfo {author} {\bibfnamefont {S.~M.}\ \bibnamefont {Girvin}},\ and\ \bibinfo {author} {\bibfnamefont {R.~J.}\ \bibnamefont {Schoelkopf}},\ }\bibfield  {title} {\bibinfo {title} {Cavity quantum electrodynamics for superconducting electrical circuits: An architecture for quantum computation},\ }\href@noop {} {\bibfield  {journal} {\bibinfo  {journal} {Physical Review A}\ }\textbf {\bibinfo {volume} {69}},\ \bibinfo {pages} {062320} (\bibinfo {year} {2004})}\BibitemShut {NoStop}%
\bibitem [{\citenamefont {Aaronson}\ and\ \citenamefont {Arkhipov}(2011)}]{aaronson2011computational}%
  \BibitemOpen
  \bibfield  {author} {\bibinfo {author} {\bibfnamefont {S.}~\bibnamefont {Aaronson}}\ and\ \bibinfo {author} {\bibfnamefont {A.}~\bibnamefont {Arkhipov}},\ }\bibfield  {title} {\bibinfo {title} {The computational complexity of linear optics},\ }in\ \href@noop {} {\emph {\bibinfo {booktitle} {Proceedings of the forty-third annual ACM symposium on Theory of computing}}}\ (\bibinfo {year} {2011})\ pp.\ \bibinfo {pages} {333--342}\BibitemShut {NoStop}%
\bibitem [{\citenamefont {Deng}\ \emph {et~al.}(2023)\citenamefont {Deng}, \citenamefont {Gu}, \citenamefont {Liu}, \citenamefont {Gong}, \citenamefont {Su}, \citenamefont {Zhang}, \citenamefont {Tang}, \citenamefont {Jia}, \citenamefont {Xu}, \citenamefont {Chen} \emph {et~al.}}]{deng2023gaussian}%
  \BibitemOpen
  \bibfield  {author} {\bibinfo {author} {\bibfnamefont {Y.-H.}\ \bibnamefont {Deng}}, \bibinfo {author} {\bibfnamefont {Y.-C.}\ \bibnamefont {Gu}}, \bibinfo {author} {\bibfnamefont {H.-L.}\ \bibnamefont {Liu}}, \bibinfo {author} {\bibfnamefont {S.-Q.}\ \bibnamefont {Gong}}, \bibinfo {author} {\bibfnamefont {H.}~\bibnamefont {Su}}, \bibinfo {author} {\bibfnamefont {Z.-J.}\ \bibnamefont {Zhang}}, \bibinfo {author} {\bibfnamefont {H.-Y.}\ \bibnamefont {Tang}}, \bibinfo {author} {\bibfnamefont {M.-H.}\ \bibnamefont {Jia}}, \bibinfo {author} {\bibfnamefont {J.-M.}\ \bibnamefont {Xu}}, \bibinfo {author} {\bibfnamefont {M.-C.}\ \bibnamefont {Chen}}, \emph {et~al.},\ }\bibfield  {title} {\bibinfo {title} {Gaussian boson sampling with pseudo-photon-number-resolving detectors and quantum computational advantage},\ }\href@noop {} {\bibfield  {journal} {\bibinfo  {journal} {Physical review letters}\ }\textbf {\bibinfo {volume} {131}},\ \bibinfo {pages} {150601} (\bibinfo {year} {2023})}\BibitemShut {NoStop}%
\bibitem [{\citenamefont {Banchi}\ \emph {et~al.}(2020)\citenamefont {Banchi}, \citenamefont {Fingerhuth}, \citenamefont {Babej}, \citenamefont {Ing},\ and\ \citenamefont {Arrazola}}]{banchi2020molecular}%
  \BibitemOpen
  \bibfield  {author} {\bibinfo {author} {\bibfnamefont {L.}~\bibnamefont {Banchi}}, \bibinfo {author} {\bibfnamefont {M.}~\bibnamefont {Fingerhuth}}, \bibinfo {author} {\bibfnamefont {T.}~\bibnamefont {Babej}}, \bibinfo {author} {\bibfnamefont {C.}~\bibnamefont {Ing}},\ and\ \bibinfo {author} {\bibfnamefont {J.~M.}\ \bibnamefont {Arrazola}},\ }\bibfield  {title} {\bibinfo {title} {Molecular docking with gaussian boson sampling},\ }\href@noop {} {\bibfield  {journal} {\bibinfo  {journal} {Science advances}\ }\textbf {\bibinfo {volume} {6}},\ \bibinfo {pages} {eaax1950} (\bibinfo {year} {2020})}\BibitemShut {NoStop}%
\bibitem [{\citenamefont {Huh}\ \emph {et~al.}(2015)\citenamefont {Huh}, \citenamefont {Guerreschi}, \citenamefont {Peropadre}, \citenamefont {McClean},\ and\ \citenamefont {Aspuru-Guzik}}]{huh2015boson}%
  \BibitemOpen
  \bibfield  {author} {\bibinfo {author} {\bibfnamefont {J.}~\bibnamefont {Huh}}, \bibinfo {author} {\bibfnamefont {G.~G.}\ \bibnamefont {Guerreschi}}, \bibinfo {author} {\bibfnamefont {B.}~\bibnamefont {Peropadre}}, \bibinfo {author} {\bibfnamefont {J.~R.}\ \bibnamefont {McClean}},\ and\ \bibinfo {author} {\bibfnamefont {A.}~\bibnamefont {Aspuru-Guzik}},\ }\bibfield  {title} {\bibinfo {title} {Boson sampling for molecular vibronic spectra},\ }\href@noop {} {\bibfield  {journal} {\bibinfo  {journal} {Nature Photonics}\ }\textbf {\bibinfo {volume} {9}},\ \bibinfo {pages} {615} (\bibinfo {year} {2015})}\BibitemShut {NoStop}%
\bibitem [{\citenamefont {Arrazola}\ and\ \citenamefont {Bromley}(2018)}]{arrazola2018using}%
  \BibitemOpen
  \bibfield  {author} {\bibinfo {author} {\bibfnamefont {J.~M.}\ \bibnamefont {Arrazola}}\ and\ \bibinfo {author} {\bibfnamefont {T.~R.}\ \bibnamefont {Bromley}},\ }\bibfield  {title} {\bibinfo {title} {Using gaussian boson sampling to find dense subgraphs},\ }\href@noop {} {\bibfield  {journal} {\bibinfo  {journal} {Physical review letters}\ }\textbf {\bibinfo {volume} {121}},\ \bibinfo {pages} {030503} (\bibinfo {year} {2018})}\BibitemShut {NoStop}%
\bibitem [{\citenamefont {Preskill}(2018)}]{preskill2018quantum}%
  \BibitemOpen
  \bibfield  {author} {\bibinfo {author} {\bibfnamefont {J.}~\bibnamefont {Preskill}},\ }\bibfield  {title} {\bibinfo {title} {Quantum computing in the nisq era and beyond},\ }\href@noop {} {\bibfield  {journal} {\bibinfo  {journal} {Quantum}\ }\textbf {\bibinfo {volume} {2}},\ \bibinfo {pages} {79} (\bibinfo {year} {2018})}\BibitemShut {NoStop}%
\bibitem [{\citenamefont {McArdle}\ \emph {et~al.}(2020)\citenamefont {McArdle}, \citenamefont {Endo}, \citenamefont {Aspuru-Guzik}, \citenamefont {Benjamin},\ and\ \citenamefont {Yuan}}]{mcardle2020quantum}%
  \BibitemOpen
  \bibfield  {author} {\bibinfo {author} {\bibfnamefont {S.}~\bibnamefont {McArdle}}, \bibinfo {author} {\bibfnamefont {S.}~\bibnamefont {Endo}}, \bibinfo {author} {\bibfnamefont {A.}~\bibnamefont {Aspuru-Guzik}}, \bibinfo {author} {\bibfnamefont {S.~C.}\ \bibnamefont {Benjamin}},\ and\ \bibinfo {author} {\bibfnamefont {X.}~\bibnamefont {Yuan}},\ }\bibfield  {title} {\bibinfo {title} {Quantum computational chemistry},\ }\href@noop {} {\bibfield  {journal} {\bibinfo  {journal} {Reviews of Modern Physics}\ }\textbf {\bibinfo {volume} {92}},\ \bibinfo {pages} {015003} (\bibinfo {year} {2020})}\BibitemShut {NoStop}%
\bibitem [{\citenamefont {McClean}\ \emph {et~al.}(2016)\citenamefont {McClean}, \citenamefont {Romero}, \citenamefont {Babbush},\ and\ \citenamefont {Aspuru-Guzik}}]{mcclean2016theory}%
  \BibitemOpen
  \bibfield  {author} {\bibinfo {author} {\bibfnamefont {J.~R.}\ \bibnamefont {McClean}}, \bibinfo {author} {\bibfnamefont {J.}~\bibnamefont {Romero}}, \bibinfo {author} {\bibfnamefont {R.}~\bibnamefont {Babbush}},\ and\ \bibinfo {author} {\bibfnamefont {A.}~\bibnamefont {Aspuru-Guzik}},\ }\bibfield  {title} {\bibinfo {title} {The theory of variational hybrid quantum-classical algorithms},\ }\href@noop {} {\bibfield  {journal} {\bibinfo  {journal} {New Journal of Physics}\ }\textbf {\bibinfo {volume} {18}},\ \bibinfo {pages} {023023} (\bibinfo {year} {2016})}\BibitemShut {NoStop}%
\bibitem [{\citenamefont {Cerezo}\ \emph {et~al.}(2021)\citenamefont {Cerezo}, \citenamefont {Arrasmith}, \citenamefont {Babbush}, \citenamefont {Benjamin}, \citenamefont {Endo}, \citenamefont {Fujii}, \citenamefont {McClean}, \citenamefont {Mitarai}, \citenamefont {Yuan}, \citenamefont {Cincio} \emph {et~al.}}]{cerezo2021variational}%
  \BibitemOpen
  \bibfield  {author} {\bibinfo {author} {\bibfnamefont {M.}~\bibnamefont {Cerezo}}, \bibinfo {author} {\bibfnamefont {A.}~\bibnamefont {Arrasmith}}, \bibinfo {author} {\bibfnamefont {R.}~\bibnamefont {Babbush}}, \bibinfo {author} {\bibfnamefont {S.~C.}\ \bibnamefont {Benjamin}}, \bibinfo {author} {\bibfnamefont {S.}~\bibnamefont {Endo}}, \bibinfo {author} {\bibfnamefont {K.}~\bibnamefont {Fujii}}, \bibinfo {author} {\bibfnamefont {J.~R.}\ \bibnamefont {McClean}}, \bibinfo {author} {\bibfnamefont {K.}~\bibnamefont {Mitarai}}, \bibinfo {author} {\bibfnamefont {X.}~\bibnamefont {Yuan}}, \bibinfo {author} {\bibfnamefont {L.}~\bibnamefont {Cincio}}, \emph {et~al.},\ }\bibfield  {title} {\bibinfo {title} {Variational quantum algorithms},\ }\href@noop {} {\bibfield  {journal} {\bibinfo  {journal} {Nature Reviews Physics}\ }\textbf {\bibinfo {volume} {3}},\ \bibinfo {pages} {625} (\bibinfo {year} {2021})}\BibitemShut {NoStop}%
\bibitem [{\citenamefont {Jozsa}\ and\ \citenamefont {Miyake}(2008)}]{jozsa2008matchgates}%
  \BibitemOpen
  \bibfield  {author} {\bibinfo {author} {\bibfnamefont {R.}~\bibnamefont {Jozsa}}\ and\ \bibinfo {author} {\bibfnamefont {A.}~\bibnamefont {Miyake}},\ }\bibfield  {title} {\bibinfo {title} {Matchgates and classical simulation of quantum circuits},\ }\href@noop {} {\bibfield  {journal} {\bibinfo  {journal} {Proceedings of the Royal Society A: Mathematical, Physical and Engineering Sciences}\ }\textbf {\bibinfo {volume} {464}},\ \bibinfo {pages} {3089} (\bibinfo {year} {2008})}\BibitemShut {NoStop}%
\bibitem [{\citenamefont {Brandas}\ and\ \citenamefont {Sabin}(2011)}]{brandas2011advances}%
  \BibitemOpen
  \bibfield  {author} {\bibinfo {author} {\bibfnamefont {E.~J.}\ \bibnamefont {Brandas}}\ and\ \bibinfo {author} {\bibfnamefont {J.~R.}\ \bibnamefont {Sabin}},\ }\href@noop {} {\emph {\bibinfo {title} {Advances in quantum chemistry}}}\ (\bibinfo  {publisher} {Academic Press},\ \bibinfo {year} {2011})\BibitemShut {NoStop}%
\bibitem [{\citenamefont {Romero}\ \emph {et~al.}(2018)\citenamefont {Romero}, \citenamefont {Babbush}, \citenamefont {McClean}, \citenamefont {Hempel}, \citenamefont {Love},\ and\ \citenamefont {Aspuru-Guzik}}]{romero2018strategies}%
  \BibitemOpen
  \bibfield  {author} {\bibinfo {author} {\bibfnamefont {J.}~\bibnamefont {Romero}}, \bibinfo {author} {\bibfnamefont {R.}~\bibnamefont {Babbush}}, \bibinfo {author} {\bibfnamefont {J.~R.}\ \bibnamefont {McClean}}, \bibinfo {author} {\bibfnamefont {C.}~\bibnamefont {Hempel}}, \bibinfo {author} {\bibfnamefont {P.~J.}\ \bibnamefont {Love}},\ and\ \bibinfo {author} {\bibfnamefont {A.}~\bibnamefont {Aspuru-Guzik}},\ }\bibfield  {title} {\bibinfo {title} {Strategies for quantum computing molecular energies using the unitary coupled cluster ansatz},\ }\href@noop {} {\bibfield  {journal} {\bibinfo  {journal} {Quantum Science and Technology}\ }\textbf {\bibinfo {volume} {4}},\ \bibinfo {pages} {014008} (\bibinfo {year} {2018})}\BibitemShut {NoStop}%
\bibitem [{\citenamefont {Anand}\ \emph {et~al.}(2022)\citenamefont {Anand}, \citenamefont {Schleich}, \citenamefont {Alperin-Lea}, \citenamefont {Jensen}, \citenamefont {Sim}, \citenamefont {D{\'\i}az-Tinoco}, \citenamefont {Kottmann}, \citenamefont {Degroote}, \citenamefont {Izmaylov},\ and\ \citenamefont {Aspuru-Guzik}}]{anand2022quantum}%
  \BibitemOpen
  \bibfield  {author} {\bibinfo {author} {\bibfnamefont {A.}~\bibnamefont {Anand}}, \bibinfo {author} {\bibfnamefont {P.}~\bibnamefont {Schleich}}, \bibinfo {author} {\bibfnamefont {S.}~\bibnamefont {Alperin-Lea}}, \bibinfo {author} {\bibfnamefont {P.~W.}\ \bibnamefont {Jensen}}, \bibinfo {author} {\bibfnamefont {S.}~\bibnamefont {Sim}}, \bibinfo {author} {\bibfnamefont {M.}~\bibnamefont {D{\'\i}az-Tinoco}}, \bibinfo {author} {\bibfnamefont {J.~S.}\ \bibnamefont {Kottmann}}, \bibinfo {author} {\bibfnamefont {M.}~\bibnamefont {Degroote}}, \bibinfo {author} {\bibfnamefont {A.~F.}\ \bibnamefont {Izmaylov}},\ and\ \bibinfo {author} {\bibfnamefont {A.}~\bibnamefont {Aspuru-Guzik}},\ }\bibfield  {title} {\bibinfo {title} {A quantum computing view on unitary coupled cluster theory},\ }\href@noop {} {\bibfield  {journal} {\bibinfo  {journal} {Chemical Society Reviews}\ } (\bibinfo {year} {2022})}\BibitemShut {NoStop}%
\bibitem [{\citenamefont {Brod}(2021)}]{brod2021bosons}%
  \BibitemOpen
  \bibfield  {author} {\bibinfo {author} {\bibfnamefont {D.~J.}\ \bibnamefont {Brod}},\ }\bibfield  {title} {\bibinfo {title} {Bosons vs. fermions--a computational complexity perspective},\ }\href@noop {} {\bibfield  {journal} {\bibinfo  {journal} {Revista Brasileira de Ensino de F{\'\i}sica}\ }\textbf {\bibinfo {volume} {43}} (\bibinfo {year} {2021})}\BibitemShut {NoStop}%
\bibitem [{\citenamefont {Reck}\ \emph {et~al.}(1994)\citenamefont {Reck}, \citenamefont {Zeilinger}, \citenamefont {Bernstein},\ and\ \citenamefont {Bertani}}]{reck1994experimental}%
  \BibitemOpen
  \bibfield  {author} {\bibinfo {author} {\bibfnamefont {M.}~\bibnamefont {Reck}}, \bibinfo {author} {\bibfnamefont {A.}~\bibnamefont {Zeilinger}}, \bibinfo {author} {\bibfnamefont {H.~J.}\ \bibnamefont {Bernstein}},\ and\ \bibinfo {author} {\bibfnamefont {P.}~\bibnamefont {Bertani}},\ }\bibfield  {title} {\bibinfo {title} {Experimental realization of any discrete unitary operator},\ }\href@noop {} {\bibfield  {journal} {\bibinfo  {journal} {Physical review letters}\ }\textbf {\bibinfo {volume} {73}},\ \bibinfo {pages} {58} (\bibinfo {year} {1994})}\BibitemShut {NoStop}%
\bibitem [{\citenamefont {Shang}\ \emph {et~al.}(2021)\citenamefont {Shang}, \citenamefont {Chen}, \citenamefont {Yuan}, \citenamefont {Lu},\ and\ \citenamefont {Pan}}]{shang2021schr}%
  \BibitemOpen
  \bibfield  {author} {\bibinfo {author} {\bibfnamefont {Z.-X.}\ \bibnamefont {Shang}}, \bibinfo {author} {\bibfnamefont {M.-C.}\ \bibnamefont {Chen}}, \bibinfo {author} {\bibfnamefont {X.}~\bibnamefont {Yuan}}, \bibinfo {author} {\bibfnamefont {C.-Y.}\ \bibnamefont {Lu}},\ and\ \bibinfo {author} {\bibfnamefont {J.-W.}\ \bibnamefont {Pan}},\ }\bibfield  {title} {\bibinfo {title} {Schr$\backslash$" odinger-heisenberg variational quantum algorithms},\ }\href@noop {} {\bibfield  {journal} {\bibinfo  {journal} {arXiv preprint arXiv:2112.07881}\ } (\bibinfo {year} {2021})}\BibitemShut {NoStop}%
\bibitem [{\citenamefont {Mizukami}\ \emph {et~al.}(2020)\citenamefont {Mizukami}, \citenamefont {Mitarai}, \citenamefont {Nakagawa}, \citenamefont {Yamamoto}, \citenamefont {Yan},\ and\ \citenamefont {Ohnishi}}]{mizukami2020orbital}%
  \BibitemOpen
  \bibfield  {author} {\bibinfo {author} {\bibfnamefont {W.}~\bibnamefont {Mizukami}}, \bibinfo {author} {\bibfnamefont {K.}~\bibnamefont {Mitarai}}, \bibinfo {author} {\bibfnamefont {Y.~O.}\ \bibnamefont {Nakagawa}}, \bibinfo {author} {\bibfnamefont {T.}~\bibnamefont {Yamamoto}}, \bibinfo {author} {\bibfnamefont {T.}~\bibnamefont {Yan}},\ and\ \bibinfo {author} {\bibfnamefont {Y.-y.}\ \bibnamefont {Ohnishi}},\ }\bibfield  {title} {\bibinfo {title} {Orbital optimized unitary coupled cluster theory for quantum computer},\ }\href@noop {} {\bibfield  {journal} {\bibinfo  {journal} {Physical Review Research}\ }\textbf {\bibinfo {volume} {2}},\ \bibinfo {pages} {033421} (\bibinfo {year} {2020})}\BibitemShut {NoStop}%
\bibitem [{\citenamefont {D'Ariano}\ \emph {et~al.}(2007)\citenamefont {D'Ariano}, \citenamefont {Maccone},\ and\ \citenamefont {Sacchi}}]{d2007homodyne}%
  \BibitemOpen
  \bibfield  {author} {\bibinfo {author} {\bibfnamefont {G.~M.}\ \bibnamefont {D'Ariano}}, \bibinfo {author} {\bibfnamefont {L.}~\bibnamefont {Maccone}},\ and\ \bibinfo {author} {\bibfnamefont {M.~F.}\ \bibnamefont {Sacchi}},\ }\bibfield  {title} {\bibinfo {title} {Homodyne tomography and the reconstruction of quantum states of light},\ }in\ \href@noop {} {\emph {\bibinfo {booktitle} {Quantum Information With Continuous Variables of Atoms and Light}}}\ (\bibinfo  {publisher} {World Scientific},\ \bibinfo {year} {2007})\ pp.\ \bibinfo {pages} {141--158}\BibitemShut {NoStop}%
\bibitem [{\citenamefont {Weedbrook}\ \emph {et~al.}(2012)\citenamefont {Weedbrook}, \citenamefont {Pirandola}, \citenamefont {Garc{\'\i}a-Patr{\'o}n}, \citenamefont {Cerf}, \citenamefont {Ralph}, \citenamefont {Shapiro},\ and\ \citenamefont {Lloyd}}]{weedbrook2012gaussian}%
  \BibitemOpen
  \bibfield  {author} {\bibinfo {author} {\bibfnamefont {C.}~\bibnamefont {Weedbrook}}, \bibinfo {author} {\bibfnamefont {S.}~\bibnamefont {Pirandola}}, \bibinfo {author} {\bibfnamefont {R.}~\bibnamefont {Garc{\'\i}a-Patr{\'o}n}}, \bibinfo {author} {\bibfnamefont {N.~J.}\ \bibnamefont {Cerf}}, \bibinfo {author} {\bibfnamefont {T.~C.}\ \bibnamefont {Ralph}}, \bibinfo {author} {\bibfnamefont {J.~H.}\ \bibnamefont {Shapiro}},\ and\ \bibinfo {author} {\bibfnamefont {S.}~\bibnamefont {Lloyd}},\ }\bibfield  {title} {\bibinfo {title} {Gaussian quantum information},\ }\href@noop {} {\bibfield  {journal} {\bibinfo  {journal} {Reviews of Modern Physics}\ }\textbf {\bibinfo {volume} {84}},\ \bibinfo {pages} {621} (\bibinfo {year} {2012})}\BibitemShut {NoStop}%
\bibitem [{\citenamefont {Chakhmakhchyan}\ and\ \citenamefont {Cerf}(2017)}]{chakhmakhchyan2017boson}%
  \BibitemOpen
  \bibfield  {author} {\bibinfo {author} {\bibfnamefont {L.}~\bibnamefont {Chakhmakhchyan}}\ and\ \bibinfo {author} {\bibfnamefont {N.~J.}\ \bibnamefont {Cerf}},\ }\bibfield  {title} {\bibinfo {title} {Boson sampling with gaussian measurements},\ }\href@noop {} {\bibfield  {journal} {\bibinfo  {journal} {Physical Review A}\ }\textbf {\bibinfo {volume} {96}},\ \bibinfo {pages} {032326} (\bibinfo {year} {2017})}\BibitemShut {NoStop}%
\bibitem [{\citenamefont {Baek}\ \emph {et~al.}(2023)\citenamefont {Baek}, \citenamefont {Hait}, \citenamefont {Shee}, \citenamefont {Leimkuhler}, \citenamefont {Huggins}, \citenamefont {Stetina}, \citenamefont {Head-Gordon},\ and\ \citenamefont {Whaley}}]{baek2023say}%
  \BibitemOpen
  \bibfield  {author} {\bibinfo {author} {\bibfnamefont {U.}~\bibnamefont {Baek}}, \bibinfo {author} {\bibfnamefont {D.}~\bibnamefont {Hait}}, \bibinfo {author} {\bibfnamefont {J.}~\bibnamefont {Shee}}, \bibinfo {author} {\bibfnamefont {O.}~\bibnamefont {Leimkuhler}}, \bibinfo {author} {\bibfnamefont {W.~J.}\ \bibnamefont {Huggins}}, \bibinfo {author} {\bibfnamefont {T.~F.}\ \bibnamefont {Stetina}}, \bibinfo {author} {\bibfnamefont {M.}~\bibnamefont {Head-Gordon}},\ and\ \bibinfo {author} {\bibfnamefont {K.~B.}\ \bibnamefont {Whaley}},\ }\bibfield  {title} {\bibinfo {title} {Say no to optimization: A nonorthogonal quantum eigensolver},\ }\href@noop {} {\bibfield  {journal} {\bibinfo  {journal} {PRX Quantum}\ }\textbf {\bibinfo {volume} {4}},\ \bibinfo {pages} {030307} (\bibinfo {year} {2023})}\BibitemShut {NoStop}%
\bibitem [{\citenamefont {Quantum}\ \emph {et~al.}(2020)\citenamefont {Quantum}, \citenamefont {Collaborators*†}, \citenamefont {Arute}, \citenamefont {Arya}, \citenamefont {Babbush}, \citenamefont {Bacon}, \citenamefont {Bardin}, \citenamefont {Barends}, \citenamefont {Boixo}, \citenamefont {Broughton}, \citenamefont {Buckley} \emph {et~al.}}]{google2020hartree}%
  \BibitemOpen
  \bibfield  {author} {\bibinfo {author} {\bibfnamefont {G.~A.}\ \bibnamefont {Quantum}}, \bibinfo {author} {\bibnamefont {Collaborators*†}}, \bibinfo {author} {\bibfnamefont {F.}~\bibnamefont {Arute}}, \bibinfo {author} {\bibfnamefont {K.}~\bibnamefont {Arya}}, \bibinfo {author} {\bibfnamefont {R.}~\bibnamefont {Babbush}}, \bibinfo {author} {\bibfnamefont {D.}~\bibnamefont {Bacon}}, \bibinfo {author} {\bibfnamefont {J.~C.}\ \bibnamefont {Bardin}}, \bibinfo {author} {\bibfnamefont {R.}~\bibnamefont {Barends}}, \bibinfo {author} {\bibfnamefont {S.}~\bibnamefont {Boixo}}, \bibinfo {author} {\bibfnamefont {M.}~\bibnamefont {Broughton}}, \bibinfo {author} {\bibfnamefont {B.~B.}\ \bibnamefont {Buckley}}, \emph {et~al.},\ }\bibfield  {title} {\bibinfo {title} {Hartree-fock on a superconducting qubit quantum computer},\ }\href@noop {} {\bibfield  {journal} {\bibinfo  {journal} {Science}\ }\textbf {\bibinfo {volume} {369}},\ \bibinfo {pages} {1084} (\bibinfo {year} {2020})}\BibitemShut {NoStop}%
\bibitem [{\citenamefont {Wang}\ \emph {et~al.}(2019)\citenamefont {Wang}, \citenamefont {Qin}, \citenamefont {Ding}, \citenamefont {Chen}, \citenamefont {Chen}, \citenamefont {You}, \citenamefont {He}, \citenamefont {Jiang}, \citenamefont {You}, \citenamefont {Wang} \emph {et~al.}}]{wang2019boson}%
  \BibitemOpen
  \bibfield  {author} {\bibinfo {author} {\bibfnamefont {H.}~\bibnamefont {Wang}}, \bibinfo {author} {\bibfnamefont {J.}~\bibnamefont {Qin}}, \bibinfo {author} {\bibfnamefont {X.}~\bibnamefont {Ding}}, \bibinfo {author} {\bibfnamefont {M.-C.}\ \bibnamefont {Chen}}, \bibinfo {author} {\bibfnamefont {S.}~\bibnamefont {Chen}}, \bibinfo {author} {\bibfnamefont {X.}~\bibnamefont {You}}, \bibinfo {author} {\bibfnamefont {Y.-M.}\ \bibnamefont {He}}, \bibinfo {author} {\bibfnamefont {X.}~\bibnamefont {Jiang}}, \bibinfo {author} {\bibfnamefont {L.}~\bibnamefont {You}}, \bibinfo {author} {\bibfnamefont {Z.}~\bibnamefont {Wang}}, \emph {et~al.},\ }\bibfield  {title} {\bibinfo {title} {Boson sampling with 20 input photons and a 60-mode interferometer in a 1 0 14-dimensional hilbert space},\ }\href@noop {} {\bibfield  {journal} {\bibinfo  {journal} {Physical review letters}\ }\textbf {\bibinfo {volume} {123}},\ \bibinfo {pages} {250503} (\bibinfo {year} {2019})}\BibitemShut {NoStop}%
\bibitem [{\citenamefont {Wu}\ \emph {et~al.}(2021)\citenamefont {Wu}, \citenamefont {Bao}, \citenamefont {Cao}, \citenamefont {Chen}, \citenamefont {Chen}, \citenamefont {Chen}, \citenamefont {Chung}, \citenamefont {Deng}, \citenamefont {Du}, \citenamefont {Fan} \emph {et~al.}}]{wu2021strong}%
  \BibitemOpen
  \bibfield  {author} {\bibinfo {author} {\bibfnamefont {Y.}~\bibnamefont {Wu}}, \bibinfo {author} {\bibfnamefont {W.-S.}\ \bibnamefont {Bao}}, \bibinfo {author} {\bibfnamefont {S.}~\bibnamefont {Cao}}, \bibinfo {author} {\bibfnamefont {F.}~\bibnamefont {Chen}}, \bibinfo {author} {\bibfnamefont {M.-C.}\ \bibnamefont {Chen}}, \bibinfo {author} {\bibfnamefont {X.}~\bibnamefont {Chen}}, \bibinfo {author} {\bibfnamefont {T.-H.}\ \bibnamefont {Chung}}, \bibinfo {author} {\bibfnamefont {H.}~\bibnamefont {Deng}}, \bibinfo {author} {\bibfnamefont {Y.}~\bibnamefont {Du}}, \bibinfo {author} {\bibfnamefont {D.}~\bibnamefont {Fan}}, \emph {et~al.},\ }\bibfield  {title} {\bibinfo {title} {Strong quantum computational advantage using a superconducting quantum processor},\ }\href@noop {} {\bibfield  {journal} {\bibinfo  {journal} {Physical review letters}\ }\textbf {\bibinfo {volume} {127}},\ \bibinfo {pages} {180501} (\bibinfo {year} {2021})}\BibitemShut {NoStop}%
\bibitem [{\citenamefont {Pino}\ \emph {et~al.}(2021)\citenamefont {Pino}, \citenamefont {Dreiling}, \citenamefont {Figgatt}, \citenamefont {Gaebler}, \citenamefont {Moses}, \citenamefont {Allman}, \citenamefont {Baldwin}, \citenamefont {Foss-Feig}, \citenamefont {Hayes}, \citenamefont {Mayer} \emph {et~al.}}]{pino2021demonstration}%
  \BibitemOpen
  \bibfield  {author} {\bibinfo {author} {\bibfnamefont {J.~M.}\ \bibnamefont {Pino}}, \bibinfo {author} {\bibfnamefont {J.~M.}\ \bibnamefont {Dreiling}}, \bibinfo {author} {\bibfnamefont {C.}~\bibnamefont {Figgatt}}, \bibinfo {author} {\bibfnamefont {J.~P.}\ \bibnamefont {Gaebler}}, \bibinfo {author} {\bibfnamefont {S.~A.}\ \bibnamefont {Moses}}, \bibinfo {author} {\bibfnamefont {M.}~\bibnamefont {Allman}}, \bibinfo {author} {\bibfnamefont {C.}~\bibnamefont {Baldwin}}, \bibinfo {author} {\bibfnamefont {M.}~\bibnamefont {Foss-Feig}}, \bibinfo {author} {\bibfnamefont {D.}~\bibnamefont {Hayes}}, \bibinfo {author} {\bibfnamefont {K.}~\bibnamefont {Mayer}}, \emph {et~al.},\ }\bibfield  {title} {\bibinfo {title} {Demonstration of the trapped-ion quantum ccd computer architecture},\ }\href@noop {} {\bibfield  {journal} {\bibinfo  {journal} {Nature}\ }\textbf {\bibinfo {volume} {592}},\ \bibinfo {pages} {209} (\bibinfo {year} {2021})}\BibitemShut {NoStop}%
\bibitem [{\citenamefont {Bluvstein}\ \emph {et~al.}(2022)\citenamefont {Bluvstein}, \citenamefont {Levine}, \citenamefont {Semeghini}, \citenamefont {Wang}, \citenamefont {Ebadi}, \citenamefont {Kalinowski}, \citenamefont {Keesling}, \citenamefont {Maskara}, \citenamefont {Pichler}, \citenamefont {Greiner} \emph {et~al.}}]{bluvstein2022quantum}%
  \BibitemOpen
  \bibfield  {author} {\bibinfo {author} {\bibfnamefont {D.}~\bibnamefont {Bluvstein}}, \bibinfo {author} {\bibfnamefont {H.}~\bibnamefont {Levine}}, \bibinfo {author} {\bibfnamefont {G.}~\bibnamefont {Semeghini}}, \bibinfo {author} {\bibfnamefont {T.~T.}\ \bibnamefont {Wang}}, \bibinfo {author} {\bibfnamefont {S.}~\bibnamefont {Ebadi}}, \bibinfo {author} {\bibfnamefont {M.}~\bibnamefont {Kalinowski}}, \bibinfo {author} {\bibfnamefont {A.}~\bibnamefont {Keesling}}, \bibinfo {author} {\bibfnamefont {N.}~\bibnamefont {Maskara}}, \bibinfo {author} {\bibfnamefont {H.}~\bibnamefont {Pichler}}, \bibinfo {author} {\bibfnamefont {M.}~\bibnamefont {Greiner}}, \emph {et~al.},\ }\bibfield  {title} {\bibinfo {title} {A quantum processor based on coherent transport of entangled atom arrays},\ }\href@noop {} {\bibfield  {journal} {\bibinfo  {journal} {Nature}\ }\textbf {\bibinfo {volume} {604}},\ \bibinfo {pages} {451} (\bibinfo {year} {2022})}\BibitemShut {NoStop}%
\bibitem [{\citenamefont {Loh~Jr}\ \emph {et~al.}(1990)\citenamefont {Loh~Jr}, \citenamefont {Gubernatis}, \citenamefont {Scalettar}, \citenamefont {White}, \citenamefont {Scalapino},\ and\ \citenamefont {Sugar}}]{loh1990sign}%
  \BibitemOpen
  \bibfield  {author} {\bibinfo {author} {\bibfnamefont {E.}~\bibnamefont {Loh~Jr}}, \bibinfo {author} {\bibfnamefont {J.}~\bibnamefont {Gubernatis}}, \bibinfo {author} {\bibfnamefont {R.}~\bibnamefont {Scalettar}}, \bibinfo {author} {\bibfnamefont {S.}~\bibnamefont {White}}, \bibinfo {author} {\bibfnamefont {D.}~\bibnamefont {Scalapino}},\ and\ \bibinfo {author} {\bibfnamefont {R.}~\bibnamefont {Sugar}},\ }\bibfield  {title} {\bibinfo {title} {Sign problem in the numerical simulation of many-electron systems},\ }\href@noop {} {\bibfield  {journal} {\bibinfo  {journal} {Physical Review B}\ }\textbf {\bibinfo {volume} {41}},\ \bibinfo {pages} {9301} (\bibinfo {year} {1990})}\BibitemShut {NoStop}%
\bibitem [{\citenamefont {Lund}\ \emph {et~al.}(2017)\citenamefont {Lund}, \citenamefont {Bremner},\ and\ \citenamefont {Ralph}}]{lund2017quantum}%
  \BibitemOpen
  \bibfield  {author} {\bibinfo {author} {\bibfnamefont {A.~P.}\ \bibnamefont {Lund}}, \bibinfo {author} {\bibfnamefont {M.~J.}\ \bibnamefont {Bremner}},\ and\ \bibinfo {author} {\bibfnamefont {T.~C.}\ \bibnamefont {Ralph}},\ }\bibfield  {title} {\bibinfo {title} {Quantum sampling problems, bosonsampling and quantum supremacy},\ }\href@noop {} {\bibfield  {journal} {\bibinfo  {journal} {npj Quantum Information}\ }\textbf {\bibinfo {volume} {3}},\ \bibinfo {pages} {15} (\bibinfo {year} {2017})}\BibitemShut {NoStop}%
\bibitem [{\citenamefont {Olivares}\ \emph {et~al.}(2019)\citenamefont {Olivares}, \citenamefont {Allevi}, \citenamefont {Caiazzo}, \citenamefont {Paris},\ and\ \citenamefont {Bondani}}]{olivares2019quantum}%
  \BibitemOpen
  \bibfield  {author} {\bibinfo {author} {\bibfnamefont {S.}~\bibnamefont {Olivares}}, \bibinfo {author} {\bibfnamefont {A.}~\bibnamefont {Allevi}}, \bibinfo {author} {\bibfnamefont {G.}~\bibnamefont {Caiazzo}}, \bibinfo {author} {\bibfnamefont {M.~G.}\ \bibnamefont {Paris}},\ and\ \bibinfo {author} {\bibfnamefont {M.}~\bibnamefont {Bondani}},\ }\bibfield  {title} {\bibinfo {title} {Quantum tomography of light states by photon-number-resolving detectors},\ }\href@noop {} {\bibfield  {journal} {\bibinfo  {journal} {New Journal of Physics}\ }\textbf {\bibinfo {volume} {21}},\ \bibinfo {pages} {103045} (\bibinfo {year} {2019})}\BibitemShut {NoStop}%
\bibitem [{\citenamefont {Huang}\ \emph {et~al.}(2020)\citenamefont {Huang}, \citenamefont {Kueng},\ and\ \citenamefont {Preskill}}]{huang2020predicting}%
  \BibitemOpen
  \bibfield  {author} {\bibinfo {author} {\bibfnamefont {H.-Y.}\ \bibnamefont {Huang}}, \bibinfo {author} {\bibfnamefont {R.}~\bibnamefont {Kueng}},\ and\ \bibinfo {author} {\bibfnamefont {J.}~\bibnamefont {Preskill}},\ }\bibfield  {title} {\bibinfo {title} {Predicting many properties of a quantum system from very few measurements},\ }\href@noop {} {\bibfield  {journal} {\bibinfo  {journal} {Nature Physics}\ }\textbf {\bibinfo {volume} {16}},\ \bibinfo {pages} {1050} (\bibinfo {year} {2020})}\BibitemShut {NoStop}%
\bibitem [{\citenamefont {Ding}\ \emph {et~al.}(2016)\citenamefont {Ding}, \citenamefont {He}, \citenamefont {Duan}, \citenamefont {Gregersen}, \citenamefont {Chen}, \citenamefont {Unsleber}, \citenamefont {Maier}, \citenamefont {Schneider}, \citenamefont {Kamp}, \citenamefont {H{\"o}fling} \emph {et~al.}}]{ding2016demand}%
  \BibitemOpen
  \bibfield  {author} {\bibinfo {author} {\bibfnamefont {X.}~\bibnamefont {Ding}}, \bibinfo {author} {\bibfnamefont {Y.}~\bibnamefont {He}}, \bibinfo {author} {\bibfnamefont {Z.-C.}\ \bibnamefont {Duan}}, \bibinfo {author} {\bibfnamefont {N.}~\bibnamefont {Gregersen}}, \bibinfo {author} {\bibfnamefont {M.-C.}\ \bibnamefont {Chen}}, \bibinfo {author} {\bibfnamefont {S.}~\bibnamefont {Unsleber}}, \bibinfo {author} {\bibfnamefont {S.}~\bibnamefont {Maier}}, \bibinfo {author} {\bibfnamefont {C.}~\bibnamefont {Schneider}}, \bibinfo {author} {\bibfnamefont {M.}~\bibnamefont {Kamp}}, \bibinfo {author} {\bibfnamefont {S.}~\bibnamefont {H{\"o}fling}}, \emph {et~al.},\ }\bibfield  {title} {\bibinfo {title} {On-demand single photons with high extraction efficiency and near-unity indistinguishability from a resonantly driven quantum dot in a micropillar},\ }\href@noop {} {\bibfield  {journal} {\bibinfo  {journal} {Physical review letters}\ }\textbf {\bibinfo {volume} {116}},\ \bibinfo {pages} {020401} (\bibinfo {year}
  {2016})}\BibitemShut {NoStop}%
\bibitem [{\citenamefont {Senellart}\ \emph {et~al.}(2017)\citenamefont {Senellart}, \citenamefont {Solomon},\ and\ \citenamefont {White}}]{senellart2017high}%
  \BibitemOpen
  \bibfield  {author} {\bibinfo {author} {\bibfnamefont {P.}~\bibnamefont {Senellart}}, \bibinfo {author} {\bibfnamefont {G.}~\bibnamefont {Solomon}},\ and\ \bibinfo {author} {\bibfnamefont {A.}~\bibnamefont {White}},\ }\bibfield  {title} {\bibinfo {title} {High-performance semiconductor quantum-dot single-photon sources},\ }\href@noop {} {\bibfield  {journal} {\bibinfo  {journal} {Nature nanotechnology}\ }\textbf {\bibinfo {volume} {12}},\ \bibinfo {pages} {1026} (\bibinfo {year} {2017})}\BibitemShut {NoStop}%
\bibitem [{\citenamefont {Zavatta}\ \emph {et~al.}(2011)\citenamefont {Zavatta}, \citenamefont {Fiur{\'a}{\v{s}}ek},\ and\ \citenamefont {Bellini}}]{zavatta2011high}%
  \BibitemOpen
  \bibfield  {author} {\bibinfo {author} {\bibfnamefont {A.}~\bibnamefont {Zavatta}}, \bibinfo {author} {\bibfnamefont {J.}~\bibnamefont {Fiur{\'a}{\v{s}}ek}},\ and\ \bibinfo {author} {\bibfnamefont {M.}~\bibnamefont {Bellini}},\ }\bibfield  {title} {\bibinfo {title} {A high-fidelity noiseless amplifier for quantum light states},\ }\href@noop {} {\bibfield  {journal} {\bibinfo  {journal} {Nature Photonics}\ }\textbf {\bibinfo {volume} {5}},\ \bibinfo {pages} {52} (\bibinfo {year} {2011})}\BibitemShut {NoStop}%
\bibitem [{\citenamefont {Raffaelli}\ \emph {et~al.}(2018)\citenamefont {Raffaelli}, \citenamefont {Ferranti}, \citenamefont {Mahler}, \citenamefont {Sibson}, \citenamefont {Kennard}, \citenamefont {Santamato}, \citenamefont {Sinclair}, \citenamefont {Bonneau}, \citenamefont {Thompson},\ and\ \citenamefont {Matthews}}]{raffaelli2018homodyne}%
  \BibitemOpen
  \bibfield  {author} {\bibinfo {author} {\bibfnamefont {F.}~\bibnamefont {Raffaelli}}, \bibinfo {author} {\bibfnamefont {G.}~\bibnamefont {Ferranti}}, \bibinfo {author} {\bibfnamefont {D.~H.}\ \bibnamefont {Mahler}}, \bibinfo {author} {\bibfnamefont {P.}~\bibnamefont {Sibson}}, \bibinfo {author} {\bibfnamefont {J.~E.}\ \bibnamefont {Kennard}}, \bibinfo {author} {\bibfnamefont {A.}~\bibnamefont {Santamato}}, \bibinfo {author} {\bibfnamefont {G.}~\bibnamefont {Sinclair}}, \bibinfo {author} {\bibfnamefont {D.}~\bibnamefont {Bonneau}}, \bibinfo {author} {\bibfnamefont {M.~G.}\ \bibnamefont {Thompson}},\ and\ \bibinfo {author} {\bibfnamefont {J.~C.}\ \bibnamefont {Matthews}},\ }\bibfield  {title} {\bibinfo {title} {A homodyne detector integrated onto a photonic chip for measuring quantum states and generating random numbers},\ }\href@noop {} {\bibfield  {journal} {\bibinfo  {journal} {Quantum Science and Technology}\ }\textbf {\bibinfo {volume} {3}},\ \bibinfo {pages} {025003} (\bibinfo {year}
  {2018})}\BibitemShut {NoStop}%
\bibitem [{\citenamefont {Cross}(2018)}]{cross2018ibm}%
  \BibitemOpen
  \bibfield  {author} {\bibinfo {author} {\bibfnamefont {A.}~\bibnamefont {Cross}},\ }\bibfield  {title} {\bibinfo {title} {The ibm q experience and qiskit open-source quantum computing software},\ }in\ \href@noop {} {\emph {\bibinfo {booktitle} {APS March meeting abstracts}}},\ Vol.\ \bibinfo {volume} {2018}\ (\bibinfo {year} {2018})\ pp.\ \bibinfo {pages} {L58--003}\BibitemShut {NoStop}%
\bibitem [{\citenamefont {McClean}\ \emph {et~al.}(2020)\citenamefont {McClean}, \citenamefont {Rubin}, \citenamefont {Sung}, \citenamefont {Kivlichan}, \citenamefont {Bonet-Monroig}, \citenamefont {Cao}, \citenamefont {Dai}, \citenamefont {Fried}, \citenamefont {Gidney}, \citenamefont {Gimby} \emph {et~al.}}]{mcclean2020openfermion}%
  \BibitemOpen
  \bibfield  {author} {\bibinfo {author} {\bibfnamefont {J.~R.}\ \bibnamefont {McClean}}, \bibinfo {author} {\bibfnamefont {N.~C.}\ \bibnamefont {Rubin}}, \bibinfo {author} {\bibfnamefont {K.~J.}\ \bibnamefont {Sung}}, \bibinfo {author} {\bibfnamefont {I.~D.}\ \bibnamefont {Kivlichan}}, \bibinfo {author} {\bibfnamefont {X.}~\bibnamefont {Bonet-Monroig}}, \bibinfo {author} {\bibfnamefont {Y.}~\bibnamefont {Cao}}, \bibinfo {author} {\bibfnamefont {C.}~\bibnamefont {Dai}}, \bibinfo {author} {\bibfnamefont {E.~S.}\ \bibnamefont {Fried}}, \bibinfo {author} {\bibfnamefont {C.}~\bibnamefont {Gidney}}, \bibinfo {author} {\bibfnamefont {B.}~\bibnamefont {Gimby}}, \emph {et~al.},\ }\bibfield  {title} {\bibinfo {title} {Openfermion: the electronic structure package for quantum computers},\ }\href@noop {} {\bibfield  {journal} {\bibinfo  {journal} {Quantum Science and Technology}\ }\textbf {\bibinfo {volume} {5}},\ \bibinfo {pages} {034014} (\bibinfo {year} {2020})}\BibitemShut {NoStop}%
\bibitem [{\citenamefont {Sun}\ \emph {et~al.}(2018)\citenamefont {Sun}, \citenamefont {Berkelbach}, \citenamefont {Blunt}, \citenamefont {Booth}, \citenamefont {Guo}, \citenamefont {Li}, \citenamefont {Liu}, \citenamefont {McClain}, \citenamefont {Sayfutyarova}, \citenamefont {Sharma} \emph {et~al.}}]{sun2018pyscf}%
  \BibitemOpen
  \bibfield  {author} {\bibinfo {author} {\bibfnamefont {Q.}~\bibnamefont {Sun}}, \bibinfo {author} {\bibfnamefont {T.~C.}\ \bibnamefont {Berkelbach}}, \bibinfo {author} {\bibfnamefont {N.~S.}\ \bibnamefont {Blunt}}, \bibinfo {author} {\bibfnamefont {G.~H.}\ \bibnamefont {Booth}}, \bibinfo {author} {\bibfnamefont {S.}~\bibnamefont {Guo}}, \bibinfo {author} {\bibfnamefont {Z.}~\bibnamefont {Li}}, \bibinfo {author} {\bibfnamefont {J.}~\bibnamefont {Liu}}, \bibinfo {author} {\bibfnamefont {J.~D.}\ \bibnamefont {McClain}}, \bibinfo {author} {\bibfnamefont {E.~R.}\ \bibnamefont {Sayfutyarova}}, \bibinfo {author} {\bibfnamefont {S.}~\bibnamefont {Sharma}}, \emph {et~al.},\ }\bibfield  {title} {\bibinfo {title} {Pyscf: the python-based simulations of chemistry framework},\ }\href@noop {} {\bibfield  {journal} {\bibinfo  {journal} {Wiley Interdisciplinary Reviews: Computational Molecular Science}\ }\textbf {\bibinfo {volume} {8}},\ \bibinfo {pages} {e1340} (\bibinfo {year} {2018})}\BibitemShut {NoStop}%
\bibitem [{\citenamefont {Gupt}\ \emph {et~al.}(2019)\citenamefont {Gupt}, \citenamefont {Izaac},\ and\ \citenamefont {Quesada}}]{gupt2019walrus}%
  \BibitemOpen
  \bibfield  {author} {\bibinfo {author} {\bibfnamefont {B.}~\bibnamefont {Gupt}}, \bibinfo {author} {\bibfnamefont {J.}~\bibnamefont {Izaac}},\ and\ \bibinfo {author} {\bibfnamefont {N.}~\bibnamefont {Quesada}},\ }\bibfield  {title} {\bibinfo {title} {The walrus: a library for the calculation of hafnians, hermite polynomials and gaussian boson sampling},\ }\href@noop {} {\bibfield  {journal} {\bibinfo  {journal} {Journal of Open Source Software}\ }\textbf {\bibinfo {volume} {4}},\ \bibinfo {pages} {1705} (\bibinfo {year} {2019})}\BibitemShut {NoStop}%
\bibitem [{sha(2023)}]{shang2023bfvqe}%
  \BibitemOpen
  \href@noop {} {\bibinfo {title} {Example codes}},\ \bibinfo {howpublished} {\url{https://github.com/ustcszx/BS-C-VQE}} (\bibinfo {year} {2023})\BibitemShut {NoStop}%
\bibitem [{\citenamefont {Paini}\ \emph {et~al.}(2021)\citenamefont {Paini}, \citenamefont {Kalev}, \citenamefont {Padilha},\ and\ \citenamefont {Ruck}}]{paini2021estimating}%
  \BibitemOpen
  \bibfield  {author} {\bibinfo {author} {\bibfnamefont {M.}~\bibnamefont {Paini}}, \bibinfo {author} {\bibfnamefont {A.}~\bibnamefont {Kalev}}, \bibinfo {author} {\bibfnamefont {D.}~\bibnamefont {Padilha}},\ and\ \bibinfo {author} {\bibfnamefont {B.}~\bibnamefont {Ruck}},\ }\bibfield  {title} {\bibinfo {title} {Estimating expectation values using approximate quantum states},\ }\href@noop {} {\bibfield  {journal} {\bibinfo  {journal} {Quantum}\ }\textbf {\bibinfo {volume} {5}},\ \bibinfo {pages} {413} (\bibinfo {year} {2021})}\BibitemShut {NoStop}%
\bibitem [{\citenamefont {Van~Kempen}\ and\ \citenamefont {Van~Vliet}(2000)}]{van2000mean}%
  \BibitemOpen
  \bibfield  {author} {\bibinfo {author} {\bibfnamefont {G.}~\bibnamefont {Van~Kempen}}\ and\ \bibinfo {author} {\bibfnamefont {L.}~\bibnamefont {Van~Vliet}},\ }\bibfield  {title} {\bibinfo {title} {Mean and variance of ratio estimators used in fluorescence ratio imaging},\ }\href@noop {} {\bibfield  {journal} {\bibinfo  {journal} {Cytometry: The Journal of the International Society for Analytical Cytology}\ }\textbf {\bibinfo {volume} {39}},\ \bibinfo {pages} {300} (\bibinfo {year} {2000})}\BibitemShut {NoStop}%
\end{thebibliography}%
\onecolumngrid
\appendix
\section{Homodyne measurement}

First, consider a single mode \cite{d2007homodyne}. An operator $H$ can be expressed in the coherent state basis:
\begin{align}\label{a1}
H=\int\frac{d^2\alpha}{\pi}Tr[HD^\dag(\alpha)]D(\alpha)=\int_0^\pi \frac{d\phi}{\pi}\int_{-\infty}^{\infty}dr\frac{\vert r\vert}{4} Tr(H e^{irX_\phi})e^{-irX_\phi}
\end{align}
where $D(\alpha)=e^{(\alpha b^\dag-\alpha^* b)}$ is the displacement operator and $\alpha=-ire^{i\phi}/2$. The quadrature operator $X_\phi$ is defined as $ X_\phi= (b^\dag e^{i\phi} + be^{-i\phi})/2$. The expectation value of $H$ under an optical state $\rho$ can then be expressed as:
\begin{align}\label{a2}
\langle H\rangle&=Tr(H\rho)\nonumber\\&
=Tr(\int \frac{d^2\alpha}{\pi} Tr(HD^\dag(\alpha))D(\alpha)\rho)\nonumber\\&
=\int_0^\pi \frac{d\phi}{\pi}\int_{-\infty}^\infty dr \frac{ |r| }{4}Tr(He^{irX_\phi})Tr(\rho e^{-irX_\phi} )\nonumber\\&
=\int_0^\pi \frac{d\phi}{\pi}\int_{-\infty}^\infty dx \int_{-\infty}^\infty dr \frac{ |r| }{4}Tr(He^{irX_\phi})p(\phi,x)e^{-irx}\nonumber\\&
=\int_0^\pi \frac{d\phi}{\pi}\int_{-\infty}^\infty dx p(\phi,x)\int_{-\infty}^\infty dr \frac{ |r| }{4}Tr(He^{ir(X_\phi-x)})\nonumber\\&
=\int_0^\pi \frac{d\phi}{\pi}\int_{-\infty}^\infty dx p(\phi,x) K(\phi,x,H)
\end{align} 
The probability $p(\phi,x)=\langle x_\phi\vert\rho\vert x_\phi\rangle$ is obtained by measuring $\langle X_\phi\vert\rho\vert X_\phi\rangle$. We see that this is a standard Monte-Carlo estimation. The kernel function $K(\phi,x,H)$ is defined as:
\begin{align}\label{a3}
K(\phi,x,H)=\int_{-\infty}^\infty dr \frac{ |r| }{4}Tr(He^{ir(X_\phi-x)})
\end{align} 
When $H=\vert n+\lambda\rangle\langle n\vert$, the kernel function has an analytic expression:
\begin{align}\label{a4}
K(\phi,x,\vert n+\lambda\rangle\langle &n\vert)= 2 e^{-i\lambda \phi} \sqrt{\frac{n !}{(n+\lambda) !}} e^{-x^2} \sum_{\nu=0}^n \frac{(-1)^\nu}{\nu !}\left(\begin{array}{c}n+\lambda \nonumber\\n-\nu\end{array}\right) \\ & \times(2\nu+\lambda+1) ! \operatorname{Re}\left[(-i)^\lambda \mathcal{D}_{-(2 \nu+\lambda+2)}(-2 i x)\right]
\end{align} 
where $\mathcal{D}_l(x)$ the parabolic cylinder function that can be easily calculated. In our algorithm, we will need the kernel functions of $\sigma_+$,$\sigma_-$, and $Z$. According to Eq. \ref{a4}, we can calculate the range of the kernel functions of $\sigma_+$ and $\sigma_-$ which are among $[-2.07317,2.07317]$ and the kernel functions of $Z$ which is among $[-2.92345,5.33333]$.

For multi-mode situations, the generalization is natural:
\begin{align}\label{a5}
\langle H\rangle & =\int_0^\pi \frac{d\phi_1 d\phi_2...d\phi_M}{\pi^M}\int_{-\infty}^\infty dx_1 d x_2...dx_M \nonumber\\ & p(\phi_1,\phi_2,...,\phi_M,x_1,x_2,...,x_M) K(\phi_1,\phi_2,...,\phi_M,x_1,x_2,...,x_M,H)
\end{align} 
where:
\begin{align}\label{a6}
K(\phi_1,\phi_2,...,&\phi_M,x_1,x_2,...,x_M,H)=\nonumber\\ &\int_{-\infty}^\infty dr_1dr_2...dr_M\Pi_{m=1}^M \frac{ \vert r_m\vert }{4}Tr(He^{ir_m(X_{\phi_m}-x_m)})
\end{align} 
We can see from Eq.\ref{a6} that the multi-mode kernel function is a simple product of the single-mode kernel function if $H$ is a product operator, which makes the complexity of calculating the kernel function scalable. Also, we don't need to calculate all values of all kernel functions in advance. We only need to do a measurement followed by a kernel function calculation. Thus, the resource for calculating the kernel function also has linear scaling with the number of measurements.

The variance of using Eq. \ref{a2} to estimate the expectation value of $H$ depends on the kernel function. When a kernel function has the range $[a,b]$, then the variance of Eq. \ref{a2} is bounded by $\frac{(b-a)^2}{4}$. Since the kernel functions of $\sigma_+$,$\sigma_-$, and $Z$ have ranges larger than $[-2,2]$ and terms we want to evaluate in the cost function obtained by the J-W mapping is global, the kernel functions of these terms can have exponentially large ranges and thus pure homodyne measurements will be non-scalable, which leads to the hybrid measurement strategy introduced in the main text. On the other hand, since $p(\phi_1,\phi_2,...,\phi_M,x_1,x_2,...,x_M)$ has no dependence on the concrete form of $H$, if all terms in a Hamiltonian are local, then the same set of samples can be used to estimate all expectation values of these terms and will greatly reduce the sampling complexity. In fact, this well-known homodyne tomography method can be seen as a natural generalization of a recently developed random measurement protocol in qubit systems \cite{paini2021estimating}.

\section{Measurement complexity}

In this section, we will estimate the measurement complexity of estimating:
\begin{align}\label{b0}
E(\vec{\alpha},\vec{\beta})=\frac{\langle \varPhi(\vec{\alpha})\vert H^{JW}_T(\vec{\beta}) \vert \varPhi(\vec{\alpha})\rangle}{\langle \varPhi(\vec{\alpha})\vert V_C^\dag(\vec{\beta})V_C(\vec{\beta})\vert \varPhi(\vec{\alpha})\rangle}
\end{align}
assuming there is no photon loss error. Before the analysis, we will introduce the estimator that will be used. 
\begin{itemize}
\item Estimator for $\frac{E[X]}{E[Y]}$: When $X$ and $Y$ are two independent random variables, the value of the ratio of their expectations $\frac{E[X]}{E[Y]}$ can be estimated by an asymptotically unbiased estimator $\frac{\overline{X}}{\overline{Y}}$ where $\overline{X}$ and $\overline{Y}$ are the averages of $X$ and $Y$. It has been shown in Ref.\cite{van2000mean} that the expectation and the variance of this estimator are:
\begin{align}
E[\frac{\overline{X}}{\overline{Y}}]&\approx \frac{E[X]}{E[Y]}+\frac{E[X]}{E[Y]^3}Var[\overline{Y}]\label{b1}\\
Var[\frac{\overline{X}}{\overline{Y}}]&\approx \frac{E[X]^2Var[\overline{Y}]+E[Y]^2Var[\overline{X}]}{E[Y]^4}\label{b2}
\end{align}
\end{itemize}
In our case, the numerator $X$ corresponds to $\langle H^{JW}_T\rangle $ short for $\langle \varPhi(\vec{\alpha})\vert H^{JW}_T(\vec{\beta}) \vert \varPhi(\vec{\alpha})\rangle$ and the denominator $Y$ corresponds to $\langle V_C^\dag V_C\rangle$ short for $\langle \varPhi_{ref}\vert V_C^\dag(\vec{\beta})V_C(\vec{\beta})\vert \varPhi_{ref} \rangle$. 
For the numerator, we assume $H^{JW}_T(\vec{\beta})$ has the form $H^{JW}_T(\vec{\beta})=\sum_{i=1}^{m_H} h_iH_i$ with $H_i$ containing $\sigma_+$, $\sigma_-$, $Z$ and $I$. For each $P_i$, we do the photon number measurements on qubits with $Z$ and $I$ and homodyne measurements on qubits with $\sigma_+$ and $\sigma_-$. Suppose each term has at most $k_h$ $\sigma_+$ and $k_h$ $\sigma_-$, since the kernel functions of $\sigma_+$ and $\sigma_-$ have ranges among $[-2.07317,2.07317]$, the remaining $Z$ and $I$ have diagonal elements -1 and 1, the whole variance is then bounded by $2.07317^{4k_h}$. If we do repeated measurements for $N_H/m_H$ times for $P_i$, the resulting variance is bounded by $m_H2.07317^{4k_h}/N_H$. Following this, the total variance of the numerator has the bound:
\begin{align}\label{b5}
Var[\overline{\langle H^{JW}_T\rangle}]\leq\sum_{i=1}^{m_H} \frac{2.07317^{4k_h}m_H h_i^2 }{N_H}
\end{align}
For the denominator, similarly, if we assume $V_C^\dag V_C=\sum_{i=1}^{m_V} g_iV_i$ and each term has at most $k_v$ $\sigma_+$ and $k_v$ $\sigma_-$. Then, if we do measurements $N_V/m_V$ times for each $P_i$, the total variance of the denominator has the bound:
\begin{align}\label{b3}
Var[\overline{\langle V_C^\dag V_C\rangle}]\leq\sum_{i=1}^{m_V} \frac{2.07317^{4k_v}m_V g_i^2 }{N_V}
\end{align}
Assume the projection ratio $\langle Q\rangle$ short for $\langle \varPhi(\vec{\alpha})\vert Q \vert \varPhi(\vec{\alpha})\rangle$ has a lower bound: $\langle Q\rangle\geq \chi $ and define $\langle H^{JW}_T\rangle_r=\langle H^{JW}_T\rangle/\langle Q\rangle$ and $\langle V_C^\dag V_C\rangle_r=\langle V_C^\dag V_C\rangle/\langle Q\rangle$. By putting Eq.\ref{b3} and Eq.\ref{b5} into Eq.\ref{b1} and Eq.\ref{b2}, we obtain that the bias of estimating $E(\vec{\alpha},\vec{\beta})$ is bounded by:
\begin{align}\label{b6}
Bias[E(\vec{\alpha},\vec{\beta})]^2&\leq\left(\frac{\langle H^{JW}_T\rangle}{\langle V_C^\dag V_C\rangle}-\frac{\overline{\langle H^{JW}_T\rangle}}{\overline{\langle V_C^\dag V_C\rangle}}\right)^2\approx\left(\frac{\langle H^{JW}_T\rangle}{\langle V_C^\dag V_C\rangle^3}Var[\overline{\langle V_C^\dag  V_C\rangle}]\right)^2\nonumber\\& =\left(\frac{E(\vec{\alpha},\vec{\beta})}{\langle V_C^\dag V_C\rangle^2} \sum_{i=1}^{m_V} \frac{2.07317^{4k_v}m_V g_i^2 }{N_V} \right)^2\nonumber\\& = \left(\frac{E(\vec{\alpha},\vec{\beta})}{\chi^2\langle V_C^\dag V_C\rangle_r^2}\frac{2.07317^{4k_v}m_V }{N_V}\sum_{i=1}^{m_V}g_i^2\right)^2 \nonumber\\&\leq \frac{2.07317^{4k_v} m_V^2  (\sum_{i=1}^{m_V}g_i^2)^2\Vert H^{JW}_T\Vert_2^2 }{\chi^4 N_V^2}
\end{align}
and the variance of estimating $E(\vec{\alpha},\vec{\beta})$ is bounded by:
\begin{align}\label{b7}
Var[E(\vec{\alpha},\vec{\beta})]&\leq\frac{\langle H^{JW}_T\rangle^2Var[\overline{\langle V_C^\dag V_C\rangle}]+\langle V_C^\dag V_C\rangle^2Var[\overline{\langle H^{JW}_T\rangle}]}{\langle V_C^\dag V_C\rangle^4}\nonumber\\&=\frac{E(\vec{\alpha},\vec{\beta})^2\sum_{i=1}^{m_V} \frac{2.07317^{4k_v}m_V g_i^2 }{N_V}+\sum_{i=1}^{m_H} \frac{2.07317^{4k_h}m_H h_i^2 }{N_H}}{\langle V_C^\dag V_C\rangle^2}\nonumber\\& =\frac{E(\vec{\alpha},\vec{\beta})^2\frac{2.07317^{4k_v}m_V }{N_V}\sum_{i=1}^{m_V}g_i^2+\frac{2.07317^{4k_h}m_H }{N_H}\sum_{i=1}^{m_H}h_i^2}{\langle V_C^\dag V_C\rangle^2}\nonumber\\&\leq \frac{2.07317^{4k_v}   m_V( \sum_{i=1}^{m_V}g_i^2 )\Vert H^{JW}_T\Vert_2^2}{\chi^2N_V}+\frac{2.07317^{4k_h}m_H\sum_{i=1}^{m_H}h_i^2}{\chi^2N_H}
\end{align}
In the above derivation, we use the fact that for HF and CISD, $\langle V_C^\dag V_C\rangle_r\geq 1$. The mean squared error is defined as the sum of Eq.\ref{b6} and Eq.\ref{b7}. From these results, we can see that to make the measurement scalable, we need $\chi$ is not exponentially small and $k_v$ and $k_h$ are restricted values, which is true for HF and CISD.

\section{Error mitigation against photon loss errors}
Consider an optical state with N photon in M modes ($M>N$), photon loss can be regarded as a photon number exchange between the system and environment, as shown below.
\begin{equation}
\vert\psi\rangle = \sum_{i=0}^N c_i \vert \psi_{N-i}\rangle\vert i\rangle_e
\end{equation}
In this equation, $\vert\psi\rangle$ represents the whole state including the system and environment. $\vert \psi_{N-i}\rangle$ represents the $N-i$ photon state component in the system with $i$ photons escaping to the environment and $c_i$ is a coefficient determined by the efficiency of the experimental system. Since the number of electrons in a molecule is fixed, we need to post-select the state $\vert \psi_{N}\rangle$. Note that $\vert \psi_{N}\rangle=d_F\vert\psi_{N,F}\rangle+...$ where $\vert\psi_{N,F}\rangle$ is the component living in the encoding Fermion space. Thus, the true expectation value of a term $H_i$ should be:
\begin{equation}
\langle\psi_{N,F}\vert H_i\vert\psi_{N,F}\rangle
\end{equation}
Given the feature of the Hamiltonian after the J-W mapping, as explained in the main text, terms of the Hamiltonian have non-diagonal $\sigma_+$ and $\sigma_-$ for certain modes and diagonal $Z$ and $I$ for other modes. Non-diagonal terms, for example, $\sigma_+\sigma_- = \vert 10\rangle\langle 01\vert$, strongly restrict the form of state that gives rise to non-zero contribution to the expectation of the term of Hamiltonian. This can be seen by an elaborate example where $H_i$ has $k$ $\sigma_-$ and $k$ $\sigma_+$. Non-zero contributions can only be given by an optical state that has $k$ photons in non-diagonal modes, and due to the reason of the electron number conservation, $N-k$ photons should be in the diagonal modes. Therefore, in the operational level, we can first estimate $\langle\phi_{N-k,T}\vert H_i\vert\phi_{N-k,T}\rangle$ where $\vert\phi_{N-k,T}\rangle$ means $N-k$ photons in diagonal modes and $T$ means the true state where there is no more than 1 photon in each mode according to the fermion encoding. In hybrid measurements, we first directly measure the photon number in diagonal modes, so, for this condition, only when $N-k$ photons are measured with no more than 1 photon in these modes, we are allowed to go to the homodyne measurement for non-diagonal modes. After this post-selection procedure, we are measuring the state $\vert\phi_{N-k}\rangle = \alpha\vert\phi_{N-k,T}\rangle+\beta \vert\phi_{N-k,W}\rangle$, which includes the state we desired and also includes the component in the wrong subspace including those with more than one photon in a non-diagonal mode and those with less than $k$ photons in all 2$k$ non-diagonal modes. Luckily, for the undesired state $\vert\phi_{N-k,W}\rangle$, it has no non-zero contributions to the expectation value and the hybrid measurements let us to estimate $\langle H_i\rangle_{raw} = \langle\phi_{N-k}\vert H_i\vert\phi_{N-k}\rangle = \vert\alpha\vert^2\langle\phi_{N-k,T}\vert H_i\vert\phi_{N-k,T}\rangle$. Thus, we only need a correction factor $1/\vert\alpha\vert^2$. Currently, $\vert\phi_{N-k,T}\rangle$ is not equal to $\vert\psi_{N,F}\rangle$ since it has other $N$-photon components. Suppose we have $\vert\psi_{N,F}\rangle= \gamma\vert\phi_{N-k,T}\rangle+\sum_i\kappa_i\vert\phi_{i\neq N-k,T}\rangle$, since other components also have zero contributions to the expectation value, we can simply add another correction factor $\vert\gamma\vert^2$. Thus, the true result is:
\begin{equation}
\langle\psi_{N,F}\vert H_i\vert\psi_{N,F}\rangle=\frac{\vert\gamma\vert^2}{\vert\alpha\vert^2} \langle\phi_{N-k}\vert H_i\vert\phi_{N-k}\rangle
\end{equation}
To get these factors, we can add an additional measurement procedure where we do pure repeated photon number measurements for many times. Suppose during these measurements, we recorded $n_1$ times of events that detected $N$ photons in all and no more than 1 photon in each mode, $n_2$ times of events that detect $N-k$ photon in diagonal modes with no more than 1 photon in these modes, and $n_3$ events among those $n_1$ ones where $N-k$ photons in the diagonal modes and $k$ photons in the non-diagonal modes. Then we have $\vert\gamma^2\vert = \frac{n_3}{n_1}$ and $\vert\alpha\vert^2 = \frac{n_3}{n_2}$ which can be used for the true expectation value estimations:
\begin{equation}
\langle\psi_{N,F}\vert H_i\vert\psi_{N,F}\rangle\approx \frac{n_3}{n_1}\frac{n_2}{n_3}\langle H_i\rangle_{raw}
\end{equation}
whose total sampling complexity can be estimated by a similar procedure as above in the measurement complexity section.

\section{Projection ratio}
\begin{figure}[htb]
\centering
\includegraphics[width=0.7\textwidth]{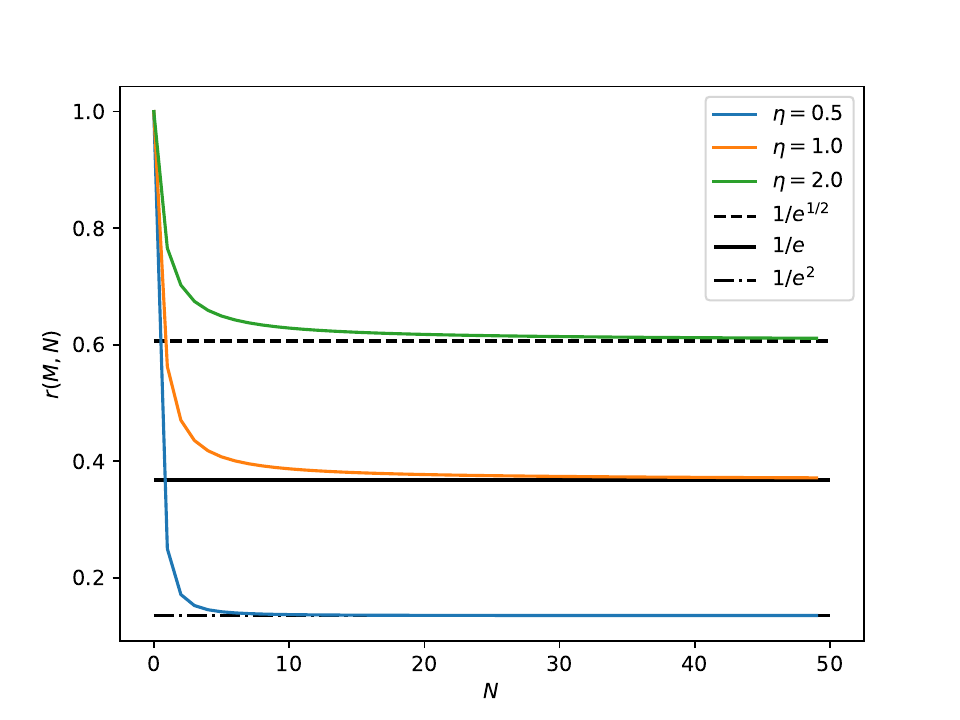}
\caption{$r(M,N)$ as functions of $N$. The number of orbitals is set to be $M=\eta N^2$ with $\eta=0.5$, $\eta=1$, and $\eta=2$.}
\label{sp1}
\end{figure}
The projection ratio highly depends on the ratio $r(M,N)=\binom{M}{N}/\binom{M+N-1}{N}$ between the dimension of Fermion Hilbert space and the dimension of Boson Hilbert space. $r(M,N)$ describes the portion of encoding space (Fermion) in physical space (Boson). In the following, we will prove that if $M=\eta N^2$, $r(M,N)$ will have a lower bound $e^{1/\eta}$. 

First, $r(M,N)$ can be re-expressed as:
\begin{align}\label{pr1}
r(M,N)=\frac{M}{M+N-1}\frac{M-1}{M+N-2}...\frac{M-N+1}{M}
\end{align}
Thus, we have:
\begin{align}\label{pr2}
r(M,N)>\left(\frac{M-N+1}{M}\right)^N
\end{align}
If $M=\eta N^2$, we have:
\begin{align}\label{pr3}
\left(\frac{M-N+1}{M}\right)^N=\frac{1}{\left(\frac{\eta N^2}{\eta N^2-N+1}\right)^N}=\frac{1}{\left(1+\frac{N-1}{\eta N^2-N+1}\right)^N}
\end{align}
The RHS of Eq.\ref{pr3} has the limit:
\begin{align}\label{pr4}
\lim_{N\rightarrow \infty}\frac{1}{\left(1+\frac{N-1}{\eta N^2-N+1}\right)^N}=\lim_{N\rightarrow \infty}\left(\frac{1}{\left(1+\frac{1}{\eta N}\right)^{\eta N}}\right)^{\frac{1}{\eta}}=e^{\frac{1}{\eta}}
\end{align}
Note that the RHS of Eq.\ref{pr3} decreases	monotonically as $N$ grows, which can be seen from Fig.\ref{sp1}. Thus, we have proved our conclusion.

\section{Molecule information in Fig.\ref{p3}}
All molecules use the STO-3G basis set. 

For the LiH molecule, we choose the active orbital $\{1, 2, 5\}$ with 2 active electrons. The orbital labels are ordered from low energy to high energy. Here, we sort the indices
of the orbitals according to their energies from low to high.

For the BeH2 molecule, the geometry is a line. We choose the active orbital $\{1, 2, 5, 6\}$ with 4 active electrons when the bond length is smaller than 1.9 and choose the active orbital $\{1, 2, 3, 6\}$ with 4 active electrons when the bond length is larger than 1.9. The orbital labels are ordered from low energy to high energy.

For the H4 molecules with line geometry and square geometry, changing of bond length in the main text means changing all H-H bonds at the same time. There are a total of 4 orbitals and 4 electrons. No reduction method is used.

Each orbital above contains two spin orbitals.

For the systemic introduction of the active space methods, see Ref.\cite{mcardle2020quantum}.

The above orbital labels can be directly used in PySCF \cite{sun2018pyscf} and Qiskit \cite{cross2018ibm} packages.
\end{document}